\documentclass[12pt,reqno]{amsart}

\usepackage{amssymb,amsxtra}
\usepackage{amsfonts}
\usepackage{amsmath} 
\usepackage{graphicx} 
\usepackage{layout}

\newcommand{\erre}{\mathbb{R}} 
\newcommand{\ci}{\mathbb{C}}
\newcommand{\natu}{\mathbb{N}} 
\newcommand{\comple}{\mathbb{C}}

 \newcommand{\al}{\alpha}
\newcommand{\ve}{\varepsilon} 
\newcommand{\la}{\lambda}

\newcommand{\ba}{\begin{eqnarray}} \newcommand{\ea}{\end{eqnarray}}
\newcommand{\be}{\begin{equation}} \newcommand{\ee}{\end{equation}}
\newcommand{\bdm}{\begin{displaymath}}
\newcommand{\edm}{\end{displaymath}} \newcommand{\brr}{\begin{array}}
\newcommand{\err}{\end{array}} 
 \newcommand{\lf}{\left}
\newcommand{\ri}{\right}

\newcommand{\bml}{\begin{gather}} \newcommand{\eml}{\end{gather}}

\newcommand{\n}{\noindent} 
\newcommand{\vs}{\vspace{0.5cm}}
\newcommand{\f}{\frac}

\newcommand{\eps}{\varepsilon}

\newtheorem{theo}{Theorem}[section]

 \newtheorem{cor}[theo]{Corollary}
\newtheorem{prop}[theo]{Proposition} 

\newcommand{\spaz}{\vspace{.5cm} \noindent}

\newenvironment{dem}{\vspace{.2cm}\noindent {\bf Proof }\\}{\newline \spaz \hspace{1cm} \hfill $\square$ \newline}

\numberwithin{equation}{section}

\begin{document}

\title[ ]{ON THE ASYMPTOTIC DYNAMICS OF A QUANTUM SYSTEM 
COMPOSED BY HEAVY AND LIGHT PARTICLES}

\author{Riccardo Adami}

\address{Adami:  Centro di Ricerca Matematica "Ennio De Giorgi"}

\curraddr{Collegio Puteano, Scuola Normale Superiore,
Piazza dei Cavalieri 4, 56100 Pisa}

\email{r.adami@sns.it}

\author{Rodolfo Figari} 
\address{ Figari: Dipartimento di Scienze Fisiche,
Universit\'a di Napoli and Sezione I.N.F.N. Napoli}

\curraddr{Via Cinthia 45, 80126 Napoli, Italy}

\email{figari@na.infn.it}

\author{Domenico Finco} 
\address{Finco: Department of Mathematics, Gakushuin University}
\curraddr{ 1-5-1 Mejiro, Toshima-ku, Tokyo
171-8588, Japan}

\email{Domenico.Finco@gakushuin.ac.jp}

\author{Alessandro Teta} \address{ Teta: Dipartimento di Matematica Pura ed
Applicata, Universit\`a di L'Aquila}

\curraddr{Via Vetoio (Coppito 1), 67010 L'Aquila, Italy}

\email{teta@univaq.it}

{\maketitle}

\vs

\begin{abstract}
We consider a non relativistic quantum system consisting of $K$ heavy and 
$N$ light particles in dimension three, where 
each heavy particle interacts with the light ones via a two-body 
potential $\alpha V$. No interaction is assumed among
particles of the same kind.

\noindent
Choosing an initial state in a product form and 
assuming $\alpha$ sufficiently small we 
characterize the asymptotic 
dynamics of the system in the limit of small mass 
ratio, with an explicit control of the error.
In the case $K=1$ the result is extended to arbitrary $\alpha$.

\n
The proof relies on a perturbative analysis 
and exploits a generalized version of the 
standard dispersive estimates for the Schr\"{o}dinger group.

\n
Exploiting the asymptotic formula, it  is also outlined an application to  the problem of 
the decoherence effect produced
on a heavy particle by the interaction with the light ones.
\end{abstract}

\vs
\vs

\section{Introduction }

\vs

\n
The study of the dynamics of a non relativistic quantum system composed by heavy and light particles is of interest in different contexts and, in particular, the search for asymptotic formulae for the wave function of the system in the small mass ratio limit is particularly relevant in many applications.

\n
In this paper we consider the case of $K$
heavy and $N$ light particles in dimension  three, where the heavy particles
interact
with the light ones via a two-body potential.
To simplify
the analysis we assume that light particles 
are not interacting among themselves
and that the  same is true for the heavy ones.

\n
We are interested in the dynamics of the system when the initial state is in a product form, i.e. no correlation among the heavy and light particles is assumed at time zero. 
Moreover we consider the regime where
only scattering processes between light
and heavy particles can occur and no other reaction channel is possible.

\n
We remark that the situation is qualitatively different from the usual case studied in 
molecular physics where the light particles, at time zero, are assumed to be in a bound state
corresponding to some energy level $ E_n (R_1, \ldots , R_K)$ produced
by the interaction potential with the heavy ones considered in the
fixed positions $R_1, \ldots , R_K$.

\n 
In that case it is well known that the standard Born-Oppenheimer approximation applies and one finds that,
for small values of the mass ratio, the
rapid motion of the light particles produces a persistent effect on the
slow (semiclassical) motion of the heavy ones, described by the effective potential $ E_n
(R_1, \ldots , R_K)$ (see e.g. \cite{h},
\cite{hj} and references therein).

\n 
The main physical motivation
at the root  of our work is the attempt to understand in a quantitative way
the loss of quantum  coherence induced on a heavy particle by the interaction
with the light ones. This  problem has attracted much interest among 
physicists in the last years
  (see
 e.g. \cite{jz}, \cite{gf}, \cite{hs}, \cite{hubhaz}, \cite{gjkksz}, \cite{bgjks}
and references therein). In particular in (\cite{hs}, \cite{hubhaz}) the
authors performed a very accurate analysis of the possible sources of collisional
decoherence in experiments  of matter wave interpherometry. We consider the
results presented in the final section  of this paper a rigorous version
of some of their results.

\n At a qualitative level, the process has been clearly described in \cite{jz}, where the starting point is the analysis of the two-body problem involving one heavy and one light particle.

\n For a small value of the mass ratio, it is reasonable to expect a separation of
two characteristic time scales, a slow one for the dynamics of the heavy
particle and a fast one for the light particle. Therefore, for an
initial state of the form $ \phi (R) \chi (r)$, where $\phi$
and $\chi$
are the initial wave functions of the heavy and the light particle
respectively, the evolution of the system is assumed to be given by the
instantaneous transition

\be
\phi(R)
\chi(r) \rightarrow \phi(R) \left( S(R) \chi \right)(r)
\label{JZ}
\ee

\n where $S(R)$ is the scattering operator corresponding to the heavy
particle fixed at the position $R$.

\n The transition (\ref{JZ}) simply means that the final state is
computed in a zero-th order adiabatic approximation, with the light
particle instantaneously scattered far away by the heavy one
considered as a fixed scattering center.

\n Notice that in (\ref{JZ}) the evolution in time of the system is completely
neglected, in the sense that time zero for the heavy particle corresponds to infinite time for the
light one.

\n 
In \cite{jz} the authors start from formula (\ref{JZ}) to 
investigate  the effect of multiple scattering events. 
They assume the existence of  collision times and a free 
dynamics of the heavy particle in between. In this way 
they restore, by hand, a time evolution of the system.

\n Our aim in this paper is to give 
a mathematical analysis of this kind of process in the more general situation of many heavy and light particles.

\n Starting from the Schr\"{o}dinger equation of the system we shall derive the asymptotic
form of the wave function for small values of the mass ratio and give an estimate
of the error.

\n The result can be considered as a rigorous derivation of 
formula (\ref{JZ}), generalized to the many particle case and modified taking into account the internal motion
of the heavy particles.

\n 
Furthermore, we shall exploit the asymptotic form
of the wave function to briefly outline how
the decoherence effect produced on the heavy
particles can be explicitely computed.

\n At this stage our analysis leaves 
untouched the question of the derivation of a master equation for the heavy particles in presence of an environment consisting of a rarefied gas of light particles (see e.g. \cite{jz}, \cite{hs}). Such derivation involves the more delicate question of the control of the limit $N \rightarrow \infty$ and requires a non trivial extension of the techniques used here.

\n The analysis presented in this paper generalizes previous results for the two-body case obtained
in \cite{dft}, where a one-dimensional system of two particles interacting via a zero-range potential was considered, and in \cite{afft}, where the result is generalized to dimension three with a generic interaction potential (se also \cite{ccf} for the case of a three-dimensional two-body system with zero-range interaction).

\vs
\n
We now give a more precise formulation of the model. Let us consider the following Hamiltonian

\be
H= \sum_{l=1}^{K}\left( -\frac{\hbar^{2}}{2M} \Delta_{R_{l}} + 
U_{l}(R_{l}) \right) + 
\sum_{j=1}^{N}\left( -\frac{\hbar^{2}}{2m}
\Delta_{r_{j}} + \alpha_{0}\sum_{l=1}^{K} V(r_{j}-R_{l}) \right)
\label{h1}
\ee

\vs
\n
acting in the Hilbert space 
${\mathcal H}= L^{2}(\erre^{3(K+N)}) =L^{2}(\erre^{3K}) \otimes 
L^{2}(\erre^{3N})$.

\n
The Hamiltonian (\ref{h1}) describes the dynamics of a quantum system composed by a sub-system of $K$ particles with position coordinates denoted by $R=(R_{1},\ldots,R_{K}) \in \erre^{3K}$, each of mass $M$ and subject to the one-body interaction potential $U_l$, plus a sub-system of $N$ particles with position coordinates denoted by $r=(r_{1},\ldots,r_{N}) 
\in \erre^{3N}$, each of mass $m$. The interaction among the particles of the two sub-systems is described by the two-body potential $\alpha_0 V$, where $\alpha_0 >0$.

\n
The potentials $U_l$, $V$ are assumed to 
be smooth and rapidly decreasing  at infinity.

\n
In order to simplify the notation we fix $\hbar=M=1$ and denote
$m= \ve$; moreover the coupling constant will be rescaled
according to $\alpha =\ve \alpha_{0} $, with $\alpha$ fixed. Then 
the Hamiltonian takes the form

\ba\label{hep}
&&H(\ve)= X +  \f{1}{\ve}
\sum_{j=1}^{N}\left( 
h_{0j} + \alpha \sum_{l=1}^{K}V(r_{j}-R_{l}) \right) 
\ea

\n
where

\ba\label{X}
&&X= \sum_{l=1}^{K}\left( -\frac{1}{2} \Delta_{R_{l}} 
+U_{l}(R_{l}) \right)
\\
&&h_{0j}=-\frac{1}{2}
\Delta_{r_{j}}
\ea

\vs
\n
We are interested in the following Cauchy problem 

\ba\label{eqSc} 
\left\{  \begin{array}{ll}
i \f{\partial }{\partial t} \Psi^{\ve}(t) = H(\ve) \Psi^{\ve}(t)
\\

\\
\Psi^{\ve}(0;R,r)
= \phi(R)\prod_{j=1}^{N}\chi_{j}(r_{j}) \equiv \phi(R) \chi(r)
\end{array}\right.
\ea

\vs

\n
where $\phi$, $\chi_{j}$ are sufficiently smooth given elements of 
$L^{2}(\erre^{3K})$ and $L^{2}(\erre^{3})$ respectively.

\n
Our aim is the characterization of the asymptotic behaviour of the solution $\Psi^{\ve}(t)$ for $\ve \rightarrow 
0$, with a control of the error.

\n
Under suitable assumptions on the potentials and the initial state, 
 we find that the asymptotic form $\Psi^{\ve}_{a}(t)$ of the wave function $\Psi^{\ve}(t)$ for $\ve \rightarrow 0$ is explicitely given by

\be\label{psia}
\Psi^{\ve}_{a}(t;R,r)= \int dR' e^{-i t X}(R,R') \phi(R') \prod_{j=1}^{N} \left( 
e^{-i \f{t}{\ve}h_{0j}} \Omega_{+}(R')^{-1} \chi_j \right)(r_j)
\ee

\n
where, for any fixed $R \in \erre^{3K}$, we have defined the following wave operator acting in the one-particle space $L^{2}(\erre^3)$ of the $j$-th light particle

\be\label{wop}
\Omega_{+}(R) \chi_j = \lim_{\tau \rightarrow + \infty} e^{i \tau h_j(R)} 
\; e^{- i \tau h_{0j}}\chi_j
\ee

\n
and in (\ref{wop}) we have denoted $h_j(R)= h_{0j} + \alpha \sum_{l=1}^{K} V(r_j - R_l)$.

\n
It should be remarked that (\ref{psia}) reduces to (\ref{JZ}) if we formally set $t=0$ and assume that $\Omega_{+}(R')^{-1} \chi_j$ can be replaced by $S(R') \chi_j$, which is approximately true for suitably chosen state $\chi_j$ (see e.g. \cite{hs}).

\n
It is important to notice that the asymptotic evolution defined by (\ref{psia}) is not factorized, due to the parametric dependence on the configuration of the heavy particles of the wave operator acting on each  light particle state.

\n
Then the asymptotic  wave function describes 
an entangled state for the whole system of heavy and light particles. In turn this implies a loss of quantum coherence for the heavy particles as a consequence of the interaction with the light ones.

\n
The precise formulation of the approximation result will be given in the next section. 
Here we only mention that in the case of an arbitrary number $K$ of heavy particles our result holds for $\alpha$ sufficiently small, while in the simpler case $K=1$ we can prove the result for any $\alpha$.

\vs
\n
The plan of the paper is the following.

\n
In section 2 we introduce some notation and formulate our main results, which are  summarized in theorems 1, 1$'$.

\n
In section 3 we give the main steps of the proof of theorems 1,1$'$.

\n
In section 4 we prove some  estimates for the unitary group generated by the Hamiltonian of the light particles $h_j(R)$, parametrically dependent on the position  $R$ of the heavy particles,  uniform with respect to the parameter $R$.

\n
In section 5 we collect some other technical lemmas concerning the unitary group generated by the Hamiltonian of the heavy particles $X$.

\n
In section 6 we briefly discuss a possible application of the asymptotic formula for the computation of the decoherence effect induced on a heavy particle.

\vs
\vs
\vs

\section{Results and notation}

\vs

\n 
Our main  result is given in theorem 1 below and concerns
the general case $K \geq 1$. In the special case $K=1$ we find
a stronger result, summarized in theorem 1'.

\n
The reason is that for the first case  
we follow and adapt to our situation the 
approach to dispersive estimates valid for small 
potentials as given in \cite{rs}, while for the second 
one we can prove the result for any $\alpha$ 
exploiting the approach to dispersive estimates 
via wave operators developed in \cite{y}.

\n
As a consequence we shall introduce  two sets 
of different assumptions on the potential  $V$ 
and on the initial state $\chi$ of the light particles.

\n
Let us denote by $W^{m,p}(\erre^{d})$, $H^{m}(\erre^{d})$ the standard Sobolev spaces and by $W^{m,p}_n(\erre^{d})$, $H^{m}_n(\erre^{d})$ the corresponding weighted Sobolev spaces, with $m,n,d \in \natu$, $1 \leq p \leq \infty$. 

\n
Then  we introduce the following assumptions
\vs

\n
(A-1)  $U_l \in W^{4, \infty}_{2} (\erre^{3})$, for $l=1, \ldots, K$; 

\vspace{0.2cm}
\n
(A-2) $\phi \in H^{4}_2(\erre^{3K})$ and $\|\phi\|_{L^2(\erre^{3K})}=1$;

\vs
\n
and, moreover, for the case $K \geq 1$

\vs
\n
(A-3) $V \in W^{4,1}(\erre^3) \cap H^4(\erre^3)$;

\vspace{0.2cm}
\n
(A-4) $\chi \in L^1(\erre^{3N}) \cap L^2(\erre^{3N})$, 
$\chi(r)= \prod_{j=1}^{N} \chi_j(r_j)$, and
$\|\chi_j\|_{L^2(\erre^{3})}=1$ for $j = 1, \dots N$. 

\vs

\n
while for the case  $K=1$ 

\vs

\n
(A-5) $V \in W^{4,\infty}_{\delta}(\erre^3)$, $\delta >5$,  and  $V \geq 0$;

\vspace{0.2cm}
\n
(A-6) $\chi \in W^{4,1}(\erre^{3N}) \cap H^4(\erre^{3N})$, 
$\chi(r)= \prod_{j=1}^{N} \chi_j(r_j)$, and
$\|\chi_j\|_{L^2(\erre^{3})}=1$ for $j = 1, \dots N$.

\vs
\n
We notice that, under  the above assumptions, 
the Hamiltonian (\ref{hep}) is self-adjoint and 
bounded from below in $\mathcal{H}$, the wave 
operator introduced in (\ref{wop}) exists and 
moreover the expression for the asymptotic 
wave function (\ref{psia}) makes sense.

\vs

\n
We now state our main result. Denoting by $\|\cdot \|$ the norm in $\mathcal{H}$, for the case $K \geq 1$ we have

\vs

\n
{\bf Theorem 1. }
{\em Let $K \geq 1$ and let us assume that  $U_l$, $\phi$, $V$,  $\chi$ satisfy assumptions} (A-1),(A-2),(A-3),(A-4); {\em moreover let us fix $T$, $0<T<\infty$, and define
\be
\alpha^{*} =  \frac{ \pi^{2/3}}{24 K} \;
 \|V\|^{-1/3}_{W^{4,1}} \;
\|V\|^{- 2/3}_{H^{4}}
\ee

\n
Then 
 for any $t \in (0,T]$ and $\alpha < \alpha^{\ast}$ we have

\be\label{th1}
\| \Psi^{\ve}(t) - \Psi^{\ve}_{a}(t) \| \leq C \, \sqrt{\frac{\ve}{t}}
\ee

\n
where  $C$ is a positive constant depending on the interaction, the initial state and $T$.}

\vs

\n
On the other hand, for the case $K=1$ we prove

\vs

\n
{\bf Theorem 1$'$}. {\em Let  $K=1$ and let us assume that $U$, $\phi$, $V$, $\chi$ satisfy
assumptions} (A-1),(A-2),(A-5),(A-6); {\em moreover let us fix $T$, $0<T<\infty$. Then for any $t \in (0,T]$ the estimate} (\ref{th1}) {\em holds, with a positive constant  
 $C'$  depending on the interaction, the initial state and $T$.}

\vs
\n
Let us briefly comment on the results stated in theorems 1, 1$'$.

\n
The estimate (\ref{th1}) clearly fails for $t \rightarrow 0$ and this fact is intrinsic in the expression of $\Psi^{\ve}_{a}(t)$, which doesn't approach $\Psi^{\ve}(0)$ for $t \rightarrow 0$.

\n
Another remark concerns the estimate of the error in (\ref{th1}), which is probably not optimal. Indeed in the simpler two-body case analysed in \cite{dft}, where the explicit form of the unitary group is available, the error found is $O(\ve)$.

\n
We also notice that the knowledge of the explicit dependence of the constant $C$ on the interaction, the initial state and $T$ is clearly interesting and  will be given during the proof. We shall find that $C$ grows with $T$, which is rather unnatural from the physical point of view and is a consequence of the specific method of the proof.  In the two-body case studied in \cite{afft} it is shown that the constant $C$ is bounded for $T$ large.

\n
Concerning the method of the proof, we observe that the approach is perturbative and it is essentially based on Duhamel's formula. The main technical ingredient for the estimates is a generalized version of the dispersive estimates for Schr\"{o}dinger groups.

\n
In fact, during the proof we shall consider the 
one-particle Hamiltonian for the $j$-th light particle $h_j(R)$, parametrically dependent on the positions $R \in \erre^{3K}$ of the heavy ones.

\n
In particular, we shall need estimates (uniform with respect to $R$) for the 
$L^{\infty}$-norm of derivatives with respect to $R$ of the unitary evolution $e^{-i \tau h_j(R)} \chi_j $.

\n
Apparently, such kind of estimates haven't been considered in the literature (see e.g. \cite{rs}, \cite{s}, \cite{y}) and then we exhibit a proof (see section 4) for $K \geq 1$ and small potential, following the approach of [RS], and also for $K=1$ and arbitrary potential, following \cite{y}.

\vs
\n
We conclude this section  collecting  some notation which will be frequently used throughout the paper.

\n
- For any $l=1, \ldots ,K$ we denote

\be
X_{0,l} = - \frac{1}{2} \Delta_{R_{l}}
\ee
and
\be
X_0 = \sum_{l=1}^{K} X_{0,l}
\ee

\n
- $U_l$, $U$ are multiplication operator by $U_l(R_l)$ and $U(R)= \sum_{l=1}^{K}U_l(R_l)$.

\n
- For any fixed $R \in \erre^{3K}$

\be\label{h}
h(R)=\sum_{j=1}^{N} h_j(R) =\sum_{j=1}^{N}\left( h_{0j} + \alpha \sum_{l=1}^{K}V(r_{j}-R_{l}) \right)
\ee

\vspace{0.2cm}
\n
denotes an operator in the Hilbert space $L^{2}(\erre^{3N})$, while $h_j(R)$ and $h_{0j}$ act in the one-particle space $L^{2}(\erre^3)$ of the $j$-th light particle.

\n
- For any $t>0$

\ba
&&\xi(t;R,r)= \phi(R) \left( e^{-i t h(R)}\chi \right) (r) = \phi(R) \prod_{j=1}^{N} \left( e^{-i t h_j(R)} \chi_j \right) (r_j) \label{xi}\\
&&\zeta^{\ve}(t;R,r)= 
\left[e^{-it X} \xi(\ve^{-1} 
t)\right](R,r) \label{zeta}
\ea

\vs
\n
defines two vectors $\xi(t)$, $\zeta^{\ve}(t) \in \mathcal{H}$.

\n
- $V_R$ is the function in $\erre^3$ defined by $V_R (x)= \sum_{l=1}^{K} V(x-R_l)$, for any fixed $R \in \erre^{3K}$.

\n
- $V_{jl}$ denotes the multiplication operator by 
$V(r_{j}-R_{l})$.

\n

\n
- $<\!\!x\!\!>$ is the multiplication 
operator by $(1+x^{2})^{1/2}$, for $x \in \erre^{d}$, $d \in 
\natu$.

\n
- $d\hat{r}_{j}= 
dr_{1},\ldots,dr_{j-1} dr_{j+1}, \ldots,dr_{N}$ and 
$d\hat{R}_{l} =dR_{1}, \ldots,dR_{l-1} dR_{l+1}, \ldots,dR_{K}$ 
denote two Lebesgue measures in $\erre^{3(N-1)}$ and $\erre^{3(K-1)}$ respectively.

\n
- The derivative of order $\gamma$ with respect to $s$-th component of 
$R_{m}$ is denoted by 
\ba
&&D^{\gamma}_{m,s} = \f{\partial^{\gamma}}{\partial 
R^{\gamma}_{m,s}}\; , \;\;\;\;\;\;\;
\gamma \in \natu , \;\;\;\; m=1,\ldots, K, \;\;\;\; s=1,2,3
\ea

\n
with  $D^1_{m,s}=D_{m,s}$.

\n
- As already mentioned, the norm in $\mathcal{H}$ is indicated by $\|\cdot\|$;   the norm in $L^p(\erre^3)$, in the Sobolev spaces $W^{m,p}(\erre^3)$, $H^{m}(\erre^3)$ and in the weighted Sobolev spaces   $W^{m,p}_{n}(\erre^3)$, $H_{n}^{m}(\erre^3)$ , $1 \leq p \leq \infty$, $m, n \in \natu$, will be denoted by $\| \cdot \|_{L^p}$, $\| \cdot \|_{W^{m,p}}$, $\| \cdot \|_{H^m}$, $\| \cdot \|_{W_{n}^{m,p}}$, $\| \cdot \|_{H_{n}^m}$ respectively.

\n
- We find convenient to introduce also a sort of slightly modified weighted Sobolev spaces, where both the weight and the derivatives concern the 
coordinates associated with only one of the heavy particles. More precisely, the weighted Sobolev space related to the $l$-th heavy particle, with indices $m,n \in \natu$, $1 \leq p \leq \infty$,  is defined as follows

\begin{eqnarray}
W^{m,p}_{l,n} (\erre^{3K}) & = & \left\{ f: \erre^{3K} \rightarrow 
\comple, \ <R_l>^n D_{l,1}^{\gamma_1} D_{l,2}^{\gamma_2} 
D_{l,3}^{\gamma_3}
f \in L^p (\erre^{3K}) \right. \nonumber \\
& & \left. {\mbox{\rm{for any} }} \ (\gamma_1, \gamma_2, \gamma_3) 
\in \natu^3, \
 \gamma_1+ \gamma_2 + \gamma_3 \leq m \right\} 
\end{eqnarray}

\noindent
The space $W^{m,p}_{l,n} (\erre^{3K})$ is a Banach space with the norm
\begin{equation}
\| f \|_{W^{m,p}_{l,n} (\erre^{3K})} = \sum_{\gamma_1=0}^m 
\sum_{\gamma_2 =0}^{m-\gamma_1}
\sum_{\gamma_3 =0}^{m-\gamma_1 -\gamma_2 } \|  <R_l>^n  
D_{l,1}^{\gamma_1} 
D_{l,2}^{\gamma_2} D_{l,3}^{\gamma_3} f \|_{L^p (\erre^{3K})}
\end{equation}
It is clear that for $f \in W^{m,p}_n (\erre^{3K})$
the quantity $\sum_{l=1}^K \| f \|_{W^{m,p}_{l,n}}$ 
defines a norm equivalent to the standard one.

\noindent
Moreover, we shall denote the space $W^{m,2}_{l,n} 
(\erre^{3K})$ by $H^m_{l,n}(\erre^{3K})$.

\n
- The operator norm of $A \;:\; E \rightarrow F$, where $E$, 
$F$ are Banach spaces, is denoted by $\|A\|_{\mathcal{L}(E,F)}$.

\n
- Finally, the symbol $c$ will denote a generic, positive, numerical constant.


\vs
\vs
\vs

\section{Proof of theorems 1,1$'$} 

\vs
\n
We give here the proof of our main result making repeated use of  some estimates which will be proved in sections 4, 5. 

\n
We start with the proof of theorem 1 and then we assume
 $\alpha < \alpha^{\ast}$. This condition guarantees the validity of a key technical ingredient, i.e. the  uniform dispersive estimate 
\be\label{stimdis}
\sup_{R}  
\left\| \left( \prod_{i=1}^n D^{\gamma_i}_{m_i,s_i} 
\right) e^{-i t h_j(R)}  \right\|_{{\mathcal L} (L^1, L^\infty)} 
\leq \frac{C_{\gamma}}{t^{3/2}} 
\ee
where 
$\gamma = \sum_{i=1}^n \gamma_i$ and 
the constant $C_{\gamma}$ is 
explicitely given (see (\ref{cgamma}) in section 4).

\n
The estimate (\ref{stimdis}) is valid for 
any string of integers $\gamma_i$ (including zero), 
$m=1,\ldots ,K$, $s=1,2,3$ and $\alpha <\alpha^*$. 

\n
In the proof we also make use of 
the following uniform $L^2$ estimate
\be\label{stimL2}
\sup_{R} \left\| \left( \prod_{i=1}^n D^{\gamma_i}_{m_i,s_i} 
\right) \prod_{k=1, k \neq j}^{N} e^{-i t h_k(R)} 
\chi_k \right\|_{L^2(\erre^{3(N-1)})} \leq \hat{C}_{\gamma}
\ee
where the constant $\hat{C}_{\gamma}$ is explicitely given too
(see (\ref{hatC}) in section 4). Notice that 
both $C_{\gamma}$ and $\hat{C}_{\gamma}$ 
are increasing with respect to $\gamma$.

\n
The proofs of (\ref{stimdis}) and (\ref{stimL2}) are postponed to section 4.

\n
The first step is to show that 
$\zeta^{\ve}(t)$ is a good approximation 
of $\Psi^{\ve}_{a}(t)$ and this is 
a direct consequence of the existence 
of the wave operator (\ref{wop}).

\n
Indeed, from (\ref{psia}) and (\ref{zeta}) we have
\ba\label{scaest}
&&\| \Psi^{\ve}_{a}(t) - \zeta^{\ve}(t) \|\nonumber\\
&&= \left( \int dR |\phi(R)|^2 \left\| 
\prod_{j=1}^{N} e^{-i \f{t}{\ve} h_{0j}}
\Omega_{+}(R)^{-1} \chi_j - \prod_{j=1}^{N} 
e^{-i \f{t}{\ve} h_j (R) }\chi_j 
\right\|^{2}_{L^{2}(\erre^{3N})} \right)^{1/2}\nonumber\\
&&\leq \sup_{R} \left\| 
\prod_{j=1}^{N} e^{-i \f{t}{\ve} h_{0j}}
\Omega_{+}(R)^{-1} \chi_j - \prod_{j=1}^{N} 
e^{-i \f{t}{\ve} h_j (R) }\chi_j 
\right\|_{L^{2}(\erre^{3N})}\nonumber\\
&&\leq \sup_{R} \sum_{n=1}^{N} \left\|
e^{-i \f{t}{\ve} h_1(R) }\chi_1 
\cdots e^{-i \f{t}{\ve} h_{n-1}(R)} \chi_{n-1}
\left(e^{-i \f{t}{\ve} h_{0n} }
\Omega_{+}(R)^{-1}\chi_n - e^{-i \f{t}
{\ve} h_n(R) }\chi_n \right) \right. \nonumber\\
&&\left. e^{-i \f{t}{\ve} h_{0n+1}} 
\Omega_{+}(R)^{-1} \chi_{n+1} \cdots 
e^{-i \f{t}{\ve} h_{0N}} 
\Omega_{+}(R)^{-1} \chi_{N} \right\|_{L^{2}(\erre^{3N})}\nonumber\\
&&\leq \sup_{R} \sum_{n=1}^{N} \left\| 
e^{i \f{t}{\ve} h_{0n}} 
e^{-i \f{t}{\ve} h_{n}(R) } 
\chi_n - \Omega_{+}(R)^{-1} \chi_n \right\|_{L^2}
\ea

\n
Let us recall that for any $\tau >0$
\ba
&&e^{i \tau h_{0n}} e^{-i \tau h_{n}(R) } 
\chi_n - \Omega_{+}(R)^{-1} \chi_n 
= i \alpha \int_{\tau}^{\infty} \! \!ds \; 
\;e^{i s h_{0n}} V_R e^{-i s h_n(R)} \chi_n
\ea

\n
Then using the dispersive estimate \eqref{stimdis} we conclude

\ba\label{scaest2}
&&\| \Psi^{\ve}_{a}(t) - 
\zeta^{\ve}(t) \| \leq \alpha \sup_{R} \left(
\|V_R\|_{L^2} \sum_{n=1}^{N} 
\int_{t/ \ve}^{\infty} \!\! ds 
\left\| e^{-i s h_n(R)} \chi_n \right\|_{L^{\infty}} \right) \nonumber\\
&&\leq \f{\sqrt{\ve}}{\sqrt{t}} \; 
C_0 \alpha K \|V\|_{L^2} \left(\sum_{n=1}^{N} 
\|\chi_n\|_{L^{1}}\right)
\ea

\vs

\n
The next and more delicate step is to show that $\zeta^{\ve}(t)$ 
approximates the solution $\Psi^{\ve}(t)$ .

\n
By a direct 
computation one has
\ba\label{eqzeta}
&&i \f{\partial}{\partial t}\zeta^{\ve}(t) 
= H(\ve) \zeta^{\ve}(t) + 
\mathcal{R}^{\ve}(t)
\ea
where
\ba\label{resto}
&&\mathcal{R}^{\ve}(t) =\f{\alpha}{\ve} \sum_{j=1}^{N}\sum_{l=1}^{K}
[e^{-itX}, V_{jl}] 
\xi(\ve^{-1}t)
\ea

\n
Using Duhamel's formula and writing
\be
[e^{-itX},V_{jl}]= 
(e^{-itX}-I)V_{jl} - V_{jl}(e^{-itX} - I) 
\ee
we have
\ba\label{pertlem}
&&\| \Psi^{\ve}(t) - \zeta^{\ve}(t) \| \leq \int_{0}^{t} ds 
\|\mathcal{R}^{\ve}(s) \|\nonumber\\
&&\leq \f{\alpha}{\ve} \sum_{l=1}^{K} \sum_{j=1}^{N} 
\int_{0}^{t} ds 
\left[
\| (e^{-i s X}-I)V_{jl} \xi(\ve^{-1}s) \| + 
\| V_{jl}(e^{-isX} - I) \xi(\ve^{-1}s)\| \right]\nonumber\\
&&= \alpha \sum_{l=1}^{K} \sum_{j=1}^{N} \int_{0}^{\ve^{-1} t} 
d \sigma \left( \mathcal{A}^{\ve}_{jl}(\sigma) + 
\mathcal{B}^{\ve}_{jl}(\sigma) \right)
\ea
where we have defined
\ba
\label{calA}
&&\mathcal{A}^{\ve}_{jl}(\sigma) = \| (e^{-i \ve \sigma X}-I)V_{jl} 
\xi(\sigma) \| \\
&& \nonumber\\
\label{calB}&&\mathcal{B}^{\ve}_{jl}(\sigma) = 
\| V_{jl}(e^{-i\ve \sigma X} - I) 
\xi(\sigma) \| 
\ea

\vs
\n
The problem is then reduced to the estimate of the two terms 
(\ref{calA}) and (\ref{calB}).

\n
The basic idea is that both terms are controlled by $e^{-i \ve \sigma 
X}-I$ for $\sigma$ small with respect to $\ve^{-1}t$ and by the 
dispersive character of the unitary group $e^{-i \sigma h(R)} $ for 
$\sigma$ of the order $\ve^{-1}t$.

\n
It turns out that such strategy is easily implemented for (\ref{calA}) 
while for (\ref{calB}) the estimate is a bit more involved.


\vs
3.a {\em Estimate of} $\mathcal{A}^{\ve}_{jl}(\sigma)$

\vs

\n
For the estimate of $\mathcal{A}^{\ve}_{jl}(\sigma)$, using 
the spectral theorem   we have
\ba
&&\label{stimaA1}
\mathcal{A}^{\ve}_{jl}(\sigma) \leq \ve \sigma \|X V_{jl} 
\xi(\sigma) \| \nonumber\\
&&\leq \ve \sigma \sum_{m=1}^{K} \| X_{0,m}(V_{jl} 
\xi(\sigma))\| + \ve \sigma  \| 
U V_{jl}\xi(\sigma) \| \nonumber\\
&&\leq \f{\ve \sigma}{2} \sum_{m=1}^{K} \sum_{s=1}^{3} 
\| D^{2}_{m,s} (V_{jl} \xi(\sigma)) \| + \ve \sigma
 \| 
U V_{jl}\xi(\sigma) \| \nonumber\\
&&\leq \f{\ve \sigma}{2} \sum_{m=1}^{K} \sum_{s=1}^{3} 
\sum_{\gamma=0}^{2} 
\left( \!\!\begin{array}{c}
2\\
\gamma
\end{array} \!\! \right)
\left\|D^{2 - \gamma}_{m,s} \left(V_{jl} 
\phi \!\! \prod_{k=1, k \neq j}^{N} \!\! 
e^{-i \sigma h_k(\cdot) }\chi_k\right) \! 
\cdot D^{\gamma}_{m,s}\left(e^{-i \sigma 
h_j(\cdot)} \chi_j\right)\right\| \nonumber\\
&& + \ve \sigma
 \left\| 
U V_{jl} \phi e^{-i \sigma h(\cdot)}\chi \right\|
\ea
where we have used the definition (\ref{xi}) and the Leibniz's rule.

\n
Using \eqref{stimdis}, the Leibniz's rule 
again and (\ref{stimL2}), we have
\ba\label{stimaA3}
&&\sum_{m=1}^K \sum_{s=1}^3\left\|D^{2 - \gamma}_{m,s} 
\left(V_{jl} \phi \!\! \prod_{k=1, k \neq j}^{N} 
\!\! e^{-i \sigma h_k(\cdot) }\chi_k\right) \! 
\cdot D^{\gamma}_{m,s}\left(e^{-i \sigma 
h_j(\cdot)} \chi_j\right)\right\| \nonumber\\
&&\leq \sum_{m=1}^K
  \sum_{s=1}^3 \sup_{R}\left\|  D^{\gamma}_{m,s} \,
e^{-i \sigma h_j(R)}\chi_j \right\|_{L^{\infty}} 
\left\|D^{2 - \gamma}_{m,s} \left(V_{jl} 
\phi \!\! \prod_{k=1, k \neq j}^{N} \!
\! e^{-i \sigma h_k(\cdot) }\chi_k\right)\right\|\nonumber\\
&&\leq 
\f{C_{\gamma}}{\sigma^{3/2}} \|\chi_j\|_{L^1} \! \!
\! \sum_{m=1}^K \sum_{s=1}^3 \sum_{\delta =0}^{2- \gamma}
\left( \!\!\!\begin{array}{c}
2\!-\! \gamma\\
\delta
\end{array} \!\!\! \right) \sup_{R}
\left\| D^{2 - \gamma - \delta}_{m,s} \!\! 
\prod_{k=1, k \neq j}^{N} \!\! e^{-i \sigma h_k(R) }\chi_k
\right\|_{L^2(\erre^{3(N-1)})} \!\! \nonumber \\
& & \cdot
\left\| D^{\delta}_{m,s}(V_{jl} \phi) 
\right\|_{L^2(\erre^{3(K+1)})}\nonumber\\
&&\leq \f{C_{\gamma}}{\sigma^{3/2}}  \hat{C}_2
\|\chi_j\|_{L^1} \; \sum_{m=1}^K 
\sum_{s=1}^3 \sum_{\delta =0}^{2- \gamma}
\left( \!\!\!\begin{array}{c}
2\!-\! \gamma\\
\delta
\end{array} \!\!\! \right)  \left\| D^{\delta}_{m,s}
(V_{jl} \phi) \right\|_{L^2(\erre^{3(K+1)})}
\ea

\n
Moreover
\ba\label{stimaA3primo}
&& \sum_{m=1}^K \sum_{s=1}^3 \sum_{\delta =0}^{2- \gamma}
    \left( \!\!\!\begin{array}{c}
  2\!-\! \gamma\\
    \delta
    \end{array} \!\!\! \right)
\left\| D^{\delta}_{m,s}(V_{jl} \phi) 
\right\|_{L^2(\erre^{3(K+1)})} \leq   4  \sum_{m=1}^K 
\sum_{s=1}^3
\sum_{\delta =0}^{2- \gamma} \sum_{\lambda = 0}^\delta
 \| D^\lambda_{m,s} V_{jl} \|_{L^2}  \| D^{\delta -\lambda}_{m,s} \phi
 \|_{L^2(\erre^{3K})} \nonumber \\
& & 
\leq  c \, \|V\|_{H^2} \|\phi\|_{H^2(\erre^{3K})}
\ea

\n
Then, using  (\ref{stimaA3primo})  in (\ref{stimaA3}), we obtain
\begin{eqnarray}
\label{stimaA5}
&& \sum_{m=1}^K \sum_{s=1}^3 \left\|D^{2 - \gamma}_{m,s} 
\left(V_{jl} \phi  \!\! 
\prod_{k=1, k \neq j}^{N} \!
\! e^{-i \sigma h_k(\cdot) }
\chi_k\right) \! \cdot 
D^{\gamma}_{m,s}\left(e^{-i \sigma 
h_j(\cdot)} \chi_j\right)\right\| \nonumber\\
&&\leq   \frac{c}{\sigma^{3/2}} 
C_\gamma \hat{C}_{2} \| \chi_j \|_{L^1} \|V\|_{H^2} \|\phi\|_{H^2(\erre^{3K})}
\end{eqnarray}

\n
Concerning the last term in (\ref{stimaA1}), 
we use  the uniform dispersive 
estimate (\ref{stimdis}) again
\begin{eqnarray}\label{stimaA2}
&& \| 
U V_{jl} \phi e^{-i \sigma h(\cdot)}\chi \| = \left(
\int \!\! \!dR \! \int \!\!\! dr_{j} 
|V(r_{j}-R_{l})|^{2} |U(R) \phi(R)|^{2} 
\! \int \! \!\!
d\hat{r}_{j} \left|\left(e^{-i \sigma h(R)}\chi 
\right)(r)\right|^{2}\right)^{1/2}\nonumber\\
&&= \left(
\int \!\!\! dR \! \int \!\!\! dr_{j} |V(r_{j}-R_{l})|^{2} |U(R) 
\phi(R)|^{2}\left|\left(e^{-i \sigma h_j(R)}
\chi_j \right)(r_j)\right|^{2}\right)^{1/2}\nonumber\\
&&= 
\sup_{R} 
\left\| e^{-i \sigma h_j(R)}\chi_j \right\|_{L^{\infty}} \! 
\|V\|_{L^{2}} \|U \phi\|_{L^{2}(\erre^{3K})}\nonumber \\
&&\leq \f{C_0}{\sigma^{3/2}} \| \chi_j\|_{L^1}
\|V\|_{L^{2}} \|U \|_{L^{\infty}(\erre^{3K})} 
\end{eqnarray}

\n
Using (\ref{stimaA5}), (\ref{stimaA2}) in (\ref{stimaA1}) we find
\begin{eqnarray}\label{stimaA4}
&&\mathcal{A}_{jl}^{\ve}(\sigma) \leq  c 
\f{\ve}{\sqrt{\sigma}} 
 \|\chi_j \|_{L^1} \!\! \left(
C_2 \, \hat{C}_{2}  \|V\|_{H^2} \| \phi \|_{H^2 
(\erre^{3K})} \! + C_0 \|V\|_{L^2}  
\| U \|_{L^{\infty}(\erre^{3K})} \! \right)\nonumber\\
&&
\end{eqnarray}
and then

\ba\label{stimaA6}
&&\alpha \sum_{l=1}^{K} \sum_{j=1}^{N} \int_{0}^{\ve^{-1}t} \!\! d \sigma \mathcal{A}_{jl}^{\ve}(\sigma) \nonumber\\
&&\leq c \sqrt{\ve} \sqrt{t} K \alpha \|V\|_{H^2} C_2 \hat{C}_2 \left( 1 + \|U\|_{L^{\infty}(\erre^{3K})}  \right)\! \left(\!\sum_{j=1}^{N} \|\chi_j\|_{L^1} \!\!\right)\|\phi\|_{H^2(\erre^{3K})}
\ea


\vs
\vs
3.b {\em Estimate of $\mathcal{B}^{\ve}_{jl}(\sigma)$}

\vs
\n
For the estimate of (\ref{calB}) we first introduce a convergence factor which makes finite the integral with respect to the variable $R_l$. In 
fact we  write

\ba\label{stimaB0}
&&\left( \mathcal{B}^{\ve}_{jl} (\sigma) \right)\nonumber\\
&&= 
\left[ \int \! \! dr_{j} \int \!\! dR_{l} 
\f{|V(r_{j}- R_{l})|^{2}}{<\!R_{l}\!>^{4}}
\int \!\! d \hat{r}_{j} \int \!\!  d\hat{R}_{l} 
\left| <\! R_{l} \! >^2 \left[ \left(e^{-i \ve \sigma X} -I \right)
\xi(\sigma) \right] \! (R,r) \right|^{2} \right]^{1/2}
\nonumber\\
&&\leq \left[ \int \!\! dx |V(x)|^2  \!\! \int \!\!  dy \frac{1}{(y^2 +1)^2} 
\sup_{r_{j}} \sup_{R_{l}} \!
\int \!\! d \hat{r}_{j}  \!\! \int \!\!  d\hat{R}_{l} 
\left| <\! R_{l} \! >^2 \left[ \left(e^{-i \ve \sigma X} \! -I \right)
\xi(\sigma) \right] \! (R,r) \right|^{2} \right]^{1/2}
\nonumber\\
&&= \pi \|V\|_{L^2} \left[
\sup_{r_{j}} \sup_{R_{l}} 
\int \!\! d \hat{r}_{j} \int \!\!  d\hat{R}_{l} 
\left| <\! R_{l} \! >^2 \left[ \left(e^{-i \ve \sigma X} -I \right)
\xi(\sigma) \right] \! (R,r) \right|^{2}  \right]^{1/2}
\nonumber\\
&&\leq \pi \|V\|_{L^2} \left[  
\sup_{r_{j}} \int \!\! d \hat{r}_{j} \int \!\!  d\hat{R}_{l} \ \sup_{R_l} 
\left| <\! R_{l} \! >^2 \left[ \left(e^{-i \ve \sigma X} -I \right)
\xi(\sigma) \right] \! (R,r) \right|^{2} \right]^{1/2}
\nonumber\\
&&\leq \pi \|V\|_{L^2}\sup_{r_j} \left[    \int d \hat{r}_j 
\left\|  (X_{0,l} +I) <\!R_{l}\!>^2  
 \left( e^{-i \ve \sigma X} -I\right) 
\xi (\sigma ;\cdot,r) \right\|^{2}_{L^2(\erre^{3K})}
 \right]^{1/2}
 \ea
where we have exploited the estimate for a.e. 
$y \in \erre^{q}$ and $x \in \erre^3$
\be\label{sup}
 |F(x,y)| \leq \left\{\int dx |\left[ (- \Delta_{x} +I) F \right] 
(x,y)|^{2} \right\}^{1/2} 
\ee
which holds for any function $F \in L^{2}( \erre^{3+q}) $, 
$q \in \natu $, such that $F(\cdot,y) \in H^{2}(\erre^{3})$ for a.e.
$y \in \erre^{q}$.

\n
Notice that the proof of (\ref{sup}) is simply 
obtained taking the Fourier transform of $F$ with respect to the variable 
$x$.

\n
It is convenient to introduce the abridged notation  
\be
\|f\|_{L^{\infty}_{j}L^2} = \sup_{r_j} \left[ \int d \hat{r}_j \|f(\cdot,r)\|^{2}_{L^2(\erre^{3K})} \right]^{1/2}
\ee
where $f \,:\, \erre^{3K} \times \erre^{3N} \rightarrow \ci$. 
Then using the formula
\be
<\! R_l \!>^2 \left( e^{-itX} -I\right) =  
\left( e^{-itX} -I\right) <\! R_l \!>^2 + \left[ R_l^2 , e^{-itX} \right]
\ee

\n
we have

\ba\label{stimaB1}
&&\left( \mathcal{B}^{\ve}_{jl} (\sigma) \right)\leq   \pi  \|V\|_{L^2} \left\|  (X_{0,l} +I) 
 \left( e^{-i \ve \sigma X} -I\right) 
 <\!R_{l}\!>^2 \xi (\sigma) \right\|
_{L^{\infty}_{j}L^2} \nonumber\\
&&+ \;  \pi \|V\|_{L^2}  \left\|  (X_{0,l} +I)
    [ R_{l}^2 , e^{-i \ve \sigma X} ] 
 \xi (\sigma) \right\|
_{L^{\infty}_{j}L^2} 
 \nonumber\\
 && \equiv (I) + (II)
\ea

\n
Writing 
\be
(X_{0,l} + I) ( e^{-i t X} -I) = (e^{-i t X}-I) 
(X_{0,l} + I) + \left[ X_{0,l} , e^{-i t X} \right]
\ee
\n
and using the spectral theorem for the estimate 
of $e^{-i \ve \sigma X}-I$, we have
\begin{eqnarray} \label{I}
& &(I) \leq  \pi \|V\|_{L^2}  
\left\|  \left( e^{-i \ve \sigma X} -I \right) 
\left( X_{0,l} + I \right)  <\! R_l \!>^2  
\xi (\sigma) \right\|_
{L^{\infty}_{j}L^2} 
\nonumber\\
&& + \pi  \|V\|_{L^2}  
\left\|  
\left[ X_{0,l} , e^{-i \ve \sigma X} 
\right] <\! R_l \!>^2 \xi(\sigma) 
\right\|_{L^{\infty}_{j}L^2}     \nonumber\\
&& \leq   \pi \ve \sigma \|V\|_{L^2}  \!
\left\|  X_{0} \left( X_{0,l} + I \right)\!  
<\! R_l \!>^2 \! \xi (\sigma) \right\|_{L^{\infty}_{j}L^2} +    \pi \ve \sigma \|V\|_{L^2}  \!
\left\|  U \! \left( X_{0,l} + I \right)\!  <\! R_l \!>^2 \! \xi (\sigma) 
\right\|_{L^{\infty}_{j}L^2}
\nonumber\\
&&+  \pi \|V\|_{L^2} 
\left\|  \left[ X_{0,l} , e^{-i \ve \sigma X} \right]   
<\! R_l \!>^2  \xi (\sigma) \right\|_{L^{\infty}_{j}L^2}\nonumber\\
&& \equiv (III) + (IV) + (V)
\ea
We decompose the term $(III)$ as follows
\begin{eqnarray}
&& (III) \leq  \pi \ve \sigma \|V\|_{L^2} 
\left\| X_{0}
<\! R_l \!>^2  \xi (\sigma) \right\|_{L^{\infty}_{j}L^2} +  \pi  \ve \sigma \|V\|_{L^2} 
\left\| X_{0} X_{0,l}   
<\! R_l \!>^2  \xi (\sigma) 
\right\|_{L^{\infty}_{j}L^2}
\nonumber\\
&& \equiv (IIIa) + (IIIb) 
\end{eqnarray}
To estimate 
$(IIIa)$ we take into account the definition (\ref{xi}) and  
the Leibniz's rule 
\ba\label{lra0}
&& (IIIa)  \leq
  c \, \ve \sigma \|V\|_{L^2}  
\sum_{m=1}^K \sum_{s=1}^3  
\left\| D_{m,s}^2 <\! R_l \!>^2 \phi 
 \prod_{k=1}^{N} e^{-i \sigma h_k (\cdot) } \chi_k
\right\|_{L^{\infty}_{j}L^2}
\nonumber\\
& & \leq   c \, \ve \sigma \|V\|_{L^2}  
\sum_{m=1}^K \sum_{s=1}^3 \sum_{\gamma=0}^{2}
\left(\! \!\!\begin{array}{c}
2 \\
\gamma
\end{array} \!\! \!\right)
\left\|  \left( D_{m,s}^\gamma e^{-i 
\sigma h_j (\cdot) } \chi_j   \right) \!\!
 \left(\!\!  D_{m,s}^{2-\gamma} <\! R_l \!>^2 \phi  \!\! 
 \prod_{k=1, k \neq j}^{N} e^{-i \sigma 
h_k (\cdot) } \chi_k \right)
\right\|_{L^{\infty}_{j}L^2} \nonumber \\
& & \leq   c \, \ve \sigma \|V\|_{L^2}  
\sum_{m=1}^K \sum_{s=1}^3 \sum_{\gamma=0}^{2}
\sum_{\lambda = 0}^{2- \gamma}
\left(\! \!\!\begin{array}{c}
2 \\
\gamma
\end{array} \!\! \!\right)
\left(\! \!\!\begin{array}{c}
2 - \gamma \\
\lambda
\end{array} \!\! \!\right)
 \nonumber\\
& & \cdot
\left\|  \left( D_{m,s}^\gamma e^{-i 
\sigma h_j (\cdot) } \chi_j \right)
 \left( D_{m,s}^{2-\gamma-\lambda} <\! R_l \!>^2 \phi 
\right)
 \left( D_{m,s}^{\lambda} 
 \prod_{k=1, k\neq j}^{N} e^{-i \sigma 
h_k (\cdot) } \chi_k \right)
\right\|_{L^{\infty}_{j}L^2}
\ea

\n
From (\ref{stimdis}) and (\ref{stimL2}) we have

\ba\label{lra}
& & (IIIa) \leq   c \, \ve \sigma \|V\|_{L^2}  
\sum_{m=1}^K \sum_{s=1}^3 \sum_{\gamma=0}^{2}
\sum_{\lambda = 0}^{2- \gamma}
\left(\! \!\!\begin{array}{c}
2 \\
\gamma
\end{array} \!\! \!\right)
\left(\! \!\!\begin{array}{c}
2 - \gamma \\
\lambda
\end{array} \!\! \!\right) 
\sup_{R} \left\|  D_{m,s}^{\gamma}  e^{-i \sigma 
h_j (R) } \chi_j \right\|_{L^\infty}
\nonumber \\
& & \cdot \left\|  D_{m,s}^{2-\gamma-\lambda} <\! R_l \!>^2 \phi 
\right\|_{L^2 (\erre^{3K})} \sup_{R}
\left\|  D_{m,s}^{\lambda}
 \prod_{k=1, k\neq j}^{N} e^{-i \sigma 
h_k (R) } \chi_k 
\right\|_{
L^2 (\erre^{3(N-1)})}
 \nonumber \\
& & \leq   c \, C_2 \hat C_2 \, 
\frac \ve {\sqrt{\sigma}}
 \|V\|_{L^2}  \| \chi_j \|_{L^1}
 \sum_{m=1}^{K} \sum_{s=1}^{3} \sum_{\delta =0}^{2} \|<\! R_l \!>^2 D^{\delta}_{m,s} \phi \|_{L^2(\erre^{3K})}
\end{eqnarray}
Analogously,
\ba\label{lrb0}
&& (IIIb) \leq
  c \, \ve \sigma \|V\|_{L^2}  
\sum_{m=1}^K \sum_{s=1}^3  \sum_{s'=1}^3  
\left\| D_{m,s}^2  D_{l,s'}^2 <\! R_l \!>^2 \phi 
 \prod_{k=1}^{N} e^{-i \sigma h_k (\cdot) } \chi_k
\right\|_{L^{\infty}_{j}L^2}
\nonumber\\
& & \leq   c \, \ve \sigma \|V\|_{L^2}  
\sum_{m=1}^K \sum_{s=1}^3 \sum_{\gamma=0}^{2}
 \sum_{\gamma'=0}^{2} 
\left(\! \!\!\begin{array}{c}
2 \\
\gamma
\end{array} \!\! \!\right) 
\left(\! \!\!\begin{array}{c}
2 \\
\gamma'
\end{array} \!\! \!\right) 
\nonumber\\
& & \cdot
\left\|  \left( D_{m,s}^\gamma  D_{l,s'}^{\gamma'} e^{-i 
\sigma h_j (\cdot) } \chi_j \right)
 \left( D_{m,s}^{2-\gamma}  D_{l,s'}^{2 - \gamma'}
 <\! R_l \!>^2 \phi 
 \prod_{k=1, k\neq j}^{N} e^{-i \sigma 
h_k (\cdot) } \chi_k \right)
\right\|_{L^{\infty}_{j}L^2}
 \nonumber \\
& & \leq   c \, \ve \sigma \|V\|_{L^2}  
\sum_{m=1}^K \sum_{s=1}^3  \sum_{s'=1}^3  
\sum_{\gamma=0}^{2}\sum_{\gamma'=0}^{2}
\sum_{\lambda = 0}^{2- \gamma}
\sum_{\lambda' = 0}^{2- \gamma'}
\left(\! \!\!\begin{array}{c}
2 \\
\gamma
\end{array} \!\! \!\right)
\left(\! \!\!\begin{array}{c}
2 - \gamma \\
\lambda
\end{array} \!\! \!\right)
\left(\! \!\!\begin{array}{c}
2 \\
\gamma'
\end{array} \!\! \!\right) 
\left(\! \!\!\begin{array}{c}
2 - \gamma' \\
\lambda'
\end{array} \!\! \!\right) 
\nonumber\\
& & \cdot
\left\|  \left( \! D_{m,s}^\gamma  D_{l,s'}^{\gamma'} e^{-i 
\sigma h_j (\cdot) } \chi_j \right)\!\! 
 \left( \! D_{m,s}^{2-\gamma-\lambda}  
D_{l,s'}^{2-\gamma'-\lambda'} 
\!<\! R_l \!>^2 \! \phi \! 
\right)\!\!
 \left(\! D_{m,s}^{\lambda}  D_{l,s'}^{\lambda'} 
 \! \!\prod_{k=1, k\neq j}^{N}\! e^{-i \sigma 
h_k (\cdot) } \chi_k \! \right)
\right\|_{L^{\infty}_{j}L^2}
\ea

\n
Using again (\ref{stimdis}), (\ref{stimL2}) we have 
\ba\label{lrb}
& &(IIIb)  \leq   c \, \ve \sigma \|V\|_{L^2}  
\sum_{m=1}^K \sum_{s=1}^3 \sum_{s'=1}^3
\sum_{\gamma=0}^{2}
\sum_{\gamma'=0}^{2}
\sum_{\lambda = 0}^{2- \gamma}\sum_{\lambda' = 0}^{2- \gamma'}
\left(\! \!\!\begin{array}{c}
2 \\
\gamma
\end{array} \!\! \!\right)
\left(\! \!\!\begin{array}{c}
2 \\
\gamma'
\end{array} \!\! \!\right)
\left(\! \!\!\begin{array}{c}
2 - \gamma \\
\lambda
\end{array} \!\! \!\right) 
\left(\! \!\!\begin{array}{c}
2 - \gamma' \\
\lambda'
\end{array} \!\! \!\right)
\nonumber \\
& & \cdot \sup_{R} 
\left\|  D_{m,s}^{\gamma}  D_{l,s'}^{\gamma'} e^{-i \sigma 
h_j (R) } \chi_j \right\|_{L^\infty}
 \left\|  D_{m,s}^{2-\gamma-\lambda} 
 D_{l,s'}^{2-\gamma'-\lambda'} <\! R_l \!>^2 \phi 
\right\|_{L^2 (\erre^{3K})}\nonumber\\
&&\cdot
\sup_{R} \left\|  D_{m,s}^{\lambda} D_{l,s'}^{\lambda'}
 \prod_{k=1, k\neq j}^{N} e^{-i \sigma 
h_k (R) } \chi_k 
\right\|_{L^2 (\erre^{3(N-1)})}
 \nonumber \\
& & \leq   c \,\frac \ve {\sqrt{\sigma}} C_4 \hat C_4 \,  \|V\|_{L^2}  \| \chi_j \|_{L^1}
\sum_{m=1}^{K} \sum_{s=1}^{3} \sum_{s' =1}^{3} \sum_{\delta =0}^{2} \sum_{\mu =0}^{2} \|<\! R_l \!>^2 D_{m,s}^{\delta} D_{l,s'}^{\mu} \phi \|_{L^2(\erre^{3K})}
\end{eqnarray}
From \eqref{lra} and \eqref{lrb} we obtain
\begin{equation}
\label{lr}
(III) \ \leq \   c \,\frac \ve {\sqrt{\sigma}} C_4 \hat C_4 \, 
 \|V\|_{L^2}  \| \chi_j \|_{L^1}
\sum_{m=1}^{K} \sum_{s=1}^{3} \sum_{s' =1}^{3} \sum_{\delta =0}^{2} \sum_{\mu =0}^{2} \|<\! R_l \!>^2 D_{m,s}^{\delta} D_{l,s'}^{\mu} \phi \|_{L^2(\erre^{3K})}
\end{equation}

\n
Following the same line, 
the estimate of $(IV)$ is easily obtained. 
In fact one has

\ba\label{IV}
&&(IV) \leq  \pi \ve \sigma  \|V\|_{L^2} 
\|U\|_{L^{\infty}(\erre^{3K})} \left\|
(X_{0,l} + I ) <\! R_l \!>^2 
\xi(\sigma) \right\|_{L^{\infty}_{j}L^2}
\nonumber\\
&&\leq c \; \f{\ve}{\sqrt{\sigma}}  
C_{2} \hat{C}_2 \|V\|_{L^2} 
\|U\|_{L^{\infty}(\erre^{3K})} 
\|\chi_j\|_{L^1} 
\sum_{s=1}^{3} \sum_{\mu =0}^{2} \|<\!R_l \!>^2 D_{l,s}^{\mu} \phi \|_{L^2(\erre^{3K})}
\ea

\vs
\n
We now consider the term $(V)$.

\n
In section 5 we shall prove  the following commutator 
estimate for any $f \in H^{1}_{l,0}(\erre^{3K})$ and $t\in [0,T]$
\ba\label{com1}
&&\| \left[ X_{0,l} , e^{-itX} \right] f 
\|_{L^2(\erre^{3K})} 
\leq t \, \tilde{C} \|f\|_{H^1_{l,0}(\erre^{3K})}
\ea
where the costant $\tilde{C}$ can be explicitely computed (see (\ref{Ctilde}) in section 5). 
Then, using (\ref{com1}), we have
\begin{eqnarray}\label{V}
& &(V) \leq  \pi \ve \sigma \, \tilde{C} \|V\|_{L^2}   
\sup_{r_j}\left[
 \int d \hat{r}_j  \| <\! R_l \!>^2 \xi(\sigma; \cdot, r) 
\|^2_{H^1_{l,0}(\erre^{3K})} \right]^{1/2} \nonumber\\
& &\leq \pi \ve \sigma \tilde{C} \|V\|_{L^2} 
\left\{ \left\| \left( e^{-i  \sigma h_j(\cdot)} 
\chi_j \right)   <\! R_l \!>^2 \phi  \!  \prod_{k=1,k \neq j}^{N} 
\! \!\left( e^{-i  \sigma h_k (\cdot)} \chi_k \right)
\! \right\|_{L^{\infty}_{j}L^2}
\right. \nonumber\\
& & +  \sum_{s=1}^{3}  \!  
\left\| 
\left( D_{l,s} e^{-i  \sigma h_j(\cdot)} \chi_j \right)\!
\!<\! R_l \!>^2 
\! \phi  
\!\! \prod_{k=1,k \neq j}^{N} \!\!\! 
\left( e^{-i  \sigma h_k (\cdot)} \chi_k \right)\!   
\right\|_{L^{\infty}_{j}L^2}\nonumber\\
&& +  \sum_{s=1}^{3}  \!  
\left\|  
(e^{-i  \sigma h_j(\cdot)} \chi_j ) 
(D_{l,s} \!<\! R_l \!>^2 \! \phi)  
\!\! \prod_{k=1,k \neq j}^{N} \!\!\! 
\left( e^{-i  \sigma h_k (\cdot)} \chi_k 
\right)\!  \right\|_{L^{\infty}_{j}L^2} \nonumber\\
& & + \left. \sum_{s=1}^{3}  
\!\left\|  
(e^{-i  \sigma h_j(\cdot)} \chi_j )  \!<\! R_l \!>^2 \! \phi    
D_{l,s}\left(  \prod_{k=1,k \neq j}^{N} \!\!\! 
\left( e^{-i  \sigma h_k (\cdot)} \chi_k \right)\! \right) 
\right\|_{L^{\infty}_{j}L^2}\right\}
\ea

\n
Exploiting the estimates (\ref{stimdis}), (\ref{stimL2}) we find
\ba\label{V2}
&&(V) \leq \pi \frac{\ve}{\sqrt{\sigma}} 
\tilde{C} C_1 \|V\|_{L^2} \|\chi_j\|_{L^1} 
\bigg[ 4 \|\! <\!R_l \!>^2 \phi \|
_{L^2(\erre^{3K})}  \nonumber\\
&&+ \sum_{s=1}^{3} 
\| D_{l,s} (<\!R_l \!>^2 \phi) \|_{L^2(\erre^{3K})} + 
3 \hat{C}_1 \|\! <\!R_l \!>^2 \phi \|_{L^2(\erre^{3K})} \bigg]
\nonumber\\
&&\leq c \frac{\ve}{\sqrt{\sigma}}   \tilde{C} C_1  
 \hat{C}_1 \|V\|_{L^2} \|\chi_j\|_{L^1} 
\sum_{s=1}^{3} \sum_{\mu =0}^{1}  \|<\!R_l \!>^2 D_{l,s}^{\mu} \phi \|_{L^2(\erre^{3K})}
\ea

\vs
\n
Let us consider the term  
$(II)$ in (\ref{stimaB1}). In section 5 
we shall prove the estimate  
for any $f \in H_{l,2}^{4}
(\erre^{3K})$ and $t \in [0,T]$
\ba\label{commR}
&& \left\| (X_{0,l} +I)  [ R_{l}^{2}, e^{-it X}]f 
\right\|_{L^{2}(\erre^{3K})} \leq t \, 
\bar{C} \|f\|_{H_{l,2}^{4}(\erre^{3K})}
\ea

\n
where the constant $\bar{C}$ can be 
explicitely computed (see (\ref{Cbar}) in section 5). 

\n
Exploiting (\ref{commR}), we have
\ba\label{II}
&&(II) \leq  \pi \ve  \sigma 
\bar{C}  \|V\|_{L^2} \sup_{r_j} \left[ \int d \hat{r}_j 
\| \xi(\sigma; \cdot,r)\|^2_{H^4_{l,2} (\erre^{3K})} 
\right]^{1/2} \nonumber\\
&&\leq \, \ve  \sigma \bar{C}  
\|V\|_{L^2} \sum_{\gamma_1 = 0}^{4}   
  \sum_{\gamma_2 = 0}^{4 - \gamma_1}   
\sum_{\gamma_3 = 0}^{4 - \gamma_1 - \gamma_2} 
\left\| <\! R_l \!>^2 D^{\gamma_1}_{l,1}  D^{\gamma_2}_{l,2}
 D^{\gamma_3}_{l,3} \left( 
\phi \prod_{k=1}^{N} (e^{-i \sigma h_k(\cdot)} 
\chi_k) \right) \right\|_{L^{\infty}_{j}L^2} 
\nonumber\\
&&\leq \, \ve  \sigma \bar{C}  
\|V\|_{L^2} \sum_{\gamma_1 = 0}^{4} 
  \sum_{\gamma_2 = 0}^{4 - \gamma_1}   
\sum_{\gamma_3 = 0}^{4 - \gamma_1 - \gamma_2} 
\sum_{\beta_1=0}^{\gamma_1}  \sum_{\beta_2=0}^{\gamma_2}
\sum_{\beta_3=0}^{\gamma_3}
\left( \!\!\!\begin{array}{c}
 \gamma_1\\
\beta_1
\end{array} \!\! \! \right)
\left( \!\!\!\begin{array}{c}
 \gamma_2\\
\beta_2
\end{array} \!\! \! \right)
\left( \!\!\!\begin{array}{c}
 \gamma_3\\
\beta_3
\end{array} \!\! \! \right)
\nonumber \\
& & \cdot \left\| 
 <\! R_l \!>^2 
\! D^{\gamma_1 - \beta_1}_{l,1}  D^{\gamma_2 - \beta_2}_{l,2}  
D^{\gamma_3 - \beta_3}_{l,3} \! 
\left( 
\! \phi \!\! \prod_{k=1, k\neq j}^{N} 
\!\! (e^{-i \sigma h_k(\cdot)} \chi_k)  
\! \right)
 \!  D^{\beta_1}_{l,1} D^{\beta_2}_{l,2} D^{\beta_3}_{l,3} (e^{-i 
\sigma h_j(\cdot)} \chi_j )
\right\|_{L^{\infty}_{j}L^2}
 \nonumber\\
& &\leq c \frac{\ve}{\sqrt{  \sigma}} 
\bar{C} C_4  \|V\|_{L^2}  \|\chi_j\|_{L^1}  
\!\sum_{\delta_1 = 0}^{4} 
\!\sum_{\delta_2 = 0}^{4 - \delta_1} \! 
\sum_{\delta_3 = 0}^{4 - \delta_1 - \delta_2}
\left\|  <\! R_l \!>^2 \! D^{\delta_1}_{l,1} D^{\delta_2}_{l,2} 
D^{\delta_3}_{l,3} 
\! \left( \! \phi \!\! \prod_{k=1, k\neq j}^{N} 
\! \!\! (e^{-i \sigma h_k(\cdot)} \chi_k)\!  
\! \! \right) \! \right\|_{L^{\infty}_{j}L^2}\nonumber\\
&&\leq c \frac{\ve}{ \sqrt{ \sigma}} 
\bar{C} C_{4}\hat{C}_{4} \|V\|_{L^2} 
\|\chi_j\|_{L^1} 
\sum_{\mu_1 =0}^{4} 
\sum_{\mu_2 =0}^{4 - \mu_1} 
\sum_{\mu_3 =0}^{4 - \mu_1 - \mu_2} 
\|<\!R_l \!>^2 D_{l,1}^{\mu_1}D_{l,2}^{\mu_2}
 D_{l,3}^{\mu_3} \phi \|_{L^2(\erre^{3K})}
\ea

\n
where we have repeatedly used the Leibniz's rule 
and estimates (\ref{stimdis}), (\ref{stimL2}).

\n
Taking into account (\ref{stimaB1}), (\ref{I}), 
(\ref{lr}), (\ref{IV}), (\ref{V2}), (\ref{II}) we obtain
\ba\label{stimaB2}
&&\mathcal{B}_{jl}^{\ve}(\sigma)\leq c 
\frac{\ve}{\sqrt{\sigma}} C_4 \hat{C}_4 
\|V\|_{L^2} \|\chi_j\|_{L^1} 
\left[ \sum_{m=1}^{K} \sum_{s=1}^{3} 
\sum_{s' =1}^{3} \sum_{\delta =0}^{2} 
\sum_{\mu =0}^{2} \|<\! R_l \!>^2 
D_{m,s}^{\delta} D_{l,s'}^{\mu} \phi \|_{L^2(\erre^{3K})}
\right. \nonumber \\
& & \left. + \bar C \sum_{\mu_1 =0}^{4} 
\sum_{\mu_2 =0}^{4 - \mu_1} \sum_{\mu_3 =0}^{4 - \mu_1 - \mu_2} 
\|<\!R_l \!>^2 D_{l,1}^{\mu_1}D_{l,2}^{\mu_2}
 D_{l,3}^{\mu_3} \phi \|_{L^2(\erre^{3K})}
\right. \nonumber \\
& & \left. + ( \tilde C + \| U \|_{L^\infty  (\erre^{3K})})
\sum_{s=1}^3 \sum_{\mu=0}^2 \|<\!R_l \!>^2 D_{l,s}^\mu \phi
\|_{L^2 (\erre^{3K})}
\right]
\ea
and then
\ba\label{stimaB3}
&&\alpha \sum_{l=1}^{K} 
\sum_{j=1}^{N} \int_{0}^{\ve^{-1}t} d \sigma  
\mathcal{B}_{jl}^{\ve}(\sigma)\nonumber\\
&& \leq c \sqrt{\ve} \sqrt{t} 
\alpha \|V\|_{L^2} C_4 \hat{C}_4 
\left(\sum_{j=1}^{N} \|\chi_j\|_{L^1} \right) \nonumber\\
& & \cdot
\left[ (1 + \tilde C + \| U \|_{L^\infty  (\erre^{3K})})
\sum_{l=1}^{K} \sum_{m=1}^{K} \sum_{s=1}^{3} 
\sum_{s' =1}^{3} \sum_{\delta =0}^{2} 
\sum_{\mu =0}^{2} \|<\! R_l \!>^2 
D_{m,s}^{\delta} D_{l,s'}^{\mu} \phi \|_{L^2(\erre^{3K})}
\right. \nonumber \\
& & \left. + \bar C \sum_{l=1}^{K}\sum_{\mu_1 =0}^{4} 
\sum_{\mu_2 =0}^{4 - \mu_1} \sum_{\mu_3 =0}^{4 - \mu_1 - \mu_2} 
\|<\!R_l \!>^2 D_{l,1}^{\mu_1}D_{l,2}^{\mu_2}
 D_{l,3}^{\mu_3} \phi \|_{L^2(\erre^{3K})}
\right]
\ea
Notice that assumption (A-2) guarantees that the norms 
in \eqref{stimaB3} involving $\phi$ are finite.

\n
Finally, using (\ref{stimaA6}), 
(\ref{stimaB3}) in (\ref{pertlem}) and taking 
into account (\ref{scaest2}) we conclude the proof of theorem 1.

\hfill$\Box$

\vs
\vs
\n
The proof of theorem 1$'$ is obtained following exactly the same line of the 
previous one with only slight modifications.
Then we shall limit ourselves to show the points to be modified.

\n
We fix $K=1$
and assume (A-1), (A-2), (A-5), (A-6); moreover we make use of the
following uniform estimates which hold for any value of $\alpha$
\ba
& &
\sup_{R} \left\| D^{\gamma_1}_{s_1} D^{\gamma_2}_{s_2} 
D^{\gamma_3}_{s_3} e^{-ith_k (R)} \chi_k \right\|_{L^\infty} 
\leq \f {B_\gamma} {t^{3/2}} \| \chi_k \|_{W^{\gamma,1}}
\label{ystimdis}\\ & &
\sup_R \left\| D^{\gamma_1}_{s_1} D^{\gamma_2}_{s_2} 
D^{\gamma_3}_{s_3}  \prod_{k=1}^N e^{-ith_k (R)} \chi_k \right\|_
{L^2 (\erre^{3N})} \leq \hat B_\gamma
\label{ystimL2}
\ea
where $\gamma = \sum_{i=1}^3 \gamma_i$ and $B_\gamma$ and
$\hat B_\gamma$ are positive constants, increasing with $\gamma$.

\n
The estimates \eqref{ystimdis}, \eqref{ystimL2} replace, in the case
$K=1$, the uniform estimates 
\eqref{stimdis}, \eqref{stimL2}, which hold for $\alpha < \alpha^*$ in the
general case $K \geq 1$, and will be proved in section 4.

\n
Proceeding for $K=1$ as in the proof of theorem 1, we see that estimate 
\eqref{scaest2} holds with $C_0$ replaced by $B_0$. Moreover we denote by
$\mathcal{A}_{j}^\ve (\sigma)$,  $\mathcal{B}_j^\ve (\sigma)$ the analogues of
\eqref{calA} and \eqref{calB} in the case $K=1$. Then it is easily seen that
\eqref{stimaA6} is replaced by 
\ba
&&
\alpha \sum_{j=1}^N \int_0^{\ve^{-1}t} d\sigma \, \mathcal{A}_j^\ve (\sigma)
\leq c \sqrt \ve \sqrt t \alpha \| V \|_{H^2} B_2 \hat B_2 \left( 1 + \| U \|_{L^\infty}
\right) \left( \sum_{j=1}^N \| \chi_j \|_{W^{2,1}} \right) \| \phi \|_{H^2}
\nonumber \\
\ea
Analogously, \eqref{stimaB3} is replaced by
\ba
&&
\alpha \sum_{j=1}^N \int_0^{\ve^{-1}t} d\sigma \, \mathcal{B}_j^\ve (\sigma)
\leq c \sqrt \ve \sqrt t \alpha \| V \|_{L^2} B_4 \hat B_4 
\left( 1 + \| U \|_{L^\infty}  + \tilde C + \bar C \right) 
\left( \sum_{j=1}^N  \| \chi_j \|_{W^{4,1}} \right) \| \phi \|_{H^4_2} 
\nonumber \\
\ea
and then the proof of theorem 1$'$ is complete.

\hfill$\Box$

\vs
\vs
\vs

\section{Uniform estimates for the unitary group $e^{-i t h(R)}$}
\vs
\noindent
In this section we shall prove some results concerning the  unitary group of 
the light particles and  its derivatives with
respect to $R$, which plays here the role of a parameter. 
In particular we shall find estimates uniform with respect to $R$.

\n
We denote the  one-particle Hamiltonian in $L^2(\erre^3)$ of the generic light particle for any fixed $R \in \erre^{3K}$ as follows

\ba
&&\hat{h}(R ) = h_0 + \alpha V_R \\
&&h_0 = - \frac{1}{2} \Delta \\ 
&& V_R(x) = \sum_{l=1}^{K} V(x- R_l), \;\;\;\; \; x \in \erre^3
\ea

\vs
\n
Moreover ${\mathcal R}_0(z)=( h_0 -z )^{-1}$ and ${\mathcal R}_R(z)=
( \hat{h}(R) -z )^{-1}$, $z \in \ci$, denote the resolvent of $h_0$ and $\hat{h}(R)$ respectively.

\noindent
Let us  first  recall some known results which will be used in the following of this section. 

\n
The potential $V$ is a Rollnik potential
if $\| V \|_{\mathcal R} < \infty$, where  $\| V \|_{\mathcal R}$ is given by

\begin{equation}
\| V \|_{\mathcal R} = \left( \int_{\erre^6} dx \, dy \frac{ | V(x)| | V(y)| }{|x-y|^2} \right)^{1/2}
\end{equation}

\vs
\n
It is well known (see e.g. th. I.4 in \cite{si}) that if $V\in L^1 \cap L^2$ then $V$ is a Rollnik potential and

\be\label{stimroll}
\|V\|_{\mathcal R} \leq c_1 \|V\|_{L^1}^{1/3} \|V\|_{L^2}^{2/3}\, , \;\;\;\;\;\; c_1 = \sqrt{3} \, (2 \pi)^{1/3}
\ee

\vs
\n
Furthermore, following the line of the proof in  \cite{si}, it is easy to see that 

\begin{equation}
\left( \int_{\erre^6} dx \, dy \frac{ | V_1(x)| | V_2(y)| }{|x-y|^2}\right)^{1/2}  \leq c_1
\left( \|V_1\|_{L^1}^{1/3} \|V_1\|_{L^2}^{2/3} \|V_2\|_{L^1}^{1/3} \|V_2\|_{L^2}^{2/3}\right)^{1/2}
\label{rollstorto}
\end{equation}

\vs
\n
The estimate (\ref{rollstorto}) is useful  in perturbation theory when one considers operators like

\begin{equation}
K(z ) = |V_1|^{1/2} {\mathcal R}_0(z ) |V_2|^{1/2}
\end{equation}

\vs
\n
where $z \in \mathbb{C}$. 
In fact the Hilbert-Schmidt norm of $K(z)$,  for    $ \Im \sqrt{z} \geq 0$, satisfies

\begin{equation}\label{HS}
\| K(z) \|_{{\mathcal{L}} (L^2, L^2)} 
\leq \| K(z) \|_{HS} \leq \frac{1}{2\pi} 
\left( \int_{\erre^6} dx \, dy \frac{ | V_1(x)| | V_2(y)| }{|x-y|^2}\right)^{1/2}
\end{equation}

\vs
\n
If the potential  $V$ belongs to $L^{3/2}$ then  $V$ is also Kato smoothing (see e.g.  \cite{y}, \cite{ky}), i.e. for any $ f \in L^2$ and $\lambda \geq 0$ we have

\begin{equation}
\sup_{\eps>0} \| |V|^{1/2} {\mathcal R}_0 ( \la \pm i \eps ) f \|_{L^2(d\la) L^2(dx)} =
\| |V|^{1/2} {\mathcal R}_0 ( \la \pm i 0 ) f \|_{L^2(d\la) L^2(dx)} \leq c \| V\|^{1/2}_{L^{3/2}} \| f\|_{L^2}
\label{ksmooth}
\end{equation}

\n

\vs
\noindent
The  potential  $V$ is a Kato potential if $\| V \|_{ \mathcal K} < \infty$, where 

\begin{equation}
\| V \|_{ \mathcal K} = \sup_{x \in \erre^3} \int_{\erre^3} dy \frac{|V(y)|}{|x-y|}
\end{equation}

\vs
\n
It is straightforward to prove 
that if $V \in L^1 \cap L^2$ then $V$ is a Kato
potential and the following estimate holds

\begin{equation}\label{stimkato}
\| V \|_{ \mathcal K} \leq c_2 \| V \|_{L^1}^{1/3} \| V\|_{L^2}^{2/3} , \;\;\;\;\; \; c_2 = 3\, \pi^{1/3}
\end{equation}

\bigskip
\n
Notice that $c_2 > c_1$.

\n
In the rest of this section we shall assume $V $ sufficiently smooth in order to guarantee the validity of (\ref{stimroll}), (\ref{ksmooth}), (\ref{stimkato}).

\n
The first result shows that, for $\alpha $ sufficiently small, 
 the usual dispersive estimate holds
uniformly with respect to  $R$.  

\begin{prop}
 Let us assume $V \in L^1 \cap L^2 $ and let 
 
 \begin{equation}\label{alfa*}
\al^{\ast}_{0} = \frac{2\pi^{2/3}}{3 K}    \,
\|V\|_{L^1}^{- 1/3} \|V\|_{L^2}^{- 2/3}
\end{equation}

\vs
\n
Then for any  $ \al < \al^{\ast}_{0}$ there exists a constant $C_0$ such that

\begin{equation}
\sup_R  \lf\| e^{-i t \hat{h}(R)}  \ri\|_{\mathcal{L}(L^1, L^{\infty})} \leq  \frac{C_0}{t^{3/2}}\label{disp}
\end{equation}
\end{prop}

\n

\begin{dem}
The proof closely follows the proof of th. 2.6 in  \cite{rs} and it is outlined  here only to highlight the uniformity with respect to $R$.
\n
Let us fix $\alpha < \alpha^{*}_{0}$. Taking into account (\ref{stimroll}) and the fact that $c_2 > c_1$ we have 

\be
 \f{1}{2\pi}  \| \alpha    V_R \|_{\mathcal R} \leq \f{\alpha K}{2 \pi} \|V\|_{\mathcal R} \leq 
\f{\alpha K}{2 \pi} c_2 \|V\|_{L^1}^{1/3} \|V\|_{L^2}^{2/3} \leq  \f{\alpha}{\alpha_{0}^{*}} <1
\ee

\n
It follows that the Born series for the boundary
value of the resolvent converges, that is for any real $ f,g \in C^{\infty}_0$ we have

\begin{equation}
<{\mathcal R}_R (\la + i0)f,g> - <{\mathcal R}_0 (\la + i0)f,g> = 
\sum_{l=1}^{\infty}(-\alpha )^l <{\mathcal R}_0 (\la + i0) \lf(  V_R {\mathcal R}_0 (\la + i0) \ri)^l f,g>
\label{born}
\end{equation}

\n
It is easily seen that the r.h.s of \eqref{born} is an absolutely convergent series which defines an element of $L^1(d\la)$ and
its norm is uniformly  bounded with respect to $R$. 

\n
Using
the spectral theorem and (\ref{born}) we have

\begin{eqnarray}
&&| <e^{-it \hat{h} (R)} f,g>| \leq
\displaystyle\sup_{L \geq 1}
\lf|\int_0^{\infty} d\la \,e^{it\la} \eta\lf(\frac{\sqrt{\la}}{L}\ri) \Im <{\mathcal R}_R (\la + i0)f,g> \ri| \nonumber\\
&&\leq \displaystyle \sum_{l=0}^{\infty} \alpha^l  \int_{\erre^6} dx_0 \, dx_{l+1} |f(x_0)| |g(x_{l+1})| 
\int_{\erre^{3l}} dx_1 \ldots dx_l \frac{ \prod_{j=1}^l | V_R(x_j )| }{ (2\pi)^{l+1}
\prod_{j=0}^l |x_j - x_{j+1}|}  \nonumber\\
&&\cdot \displaystyle
 \displaystyle\sup_{L \geq 1}
\lf|\int_0^{\infty} d\la \,e^{it\la} \eta\lf(\frac{\sqrt{\la}}{L}\ri)
\sin \lf( \sqrt{\la} \sum_{j=0}^{l} |x_j - x_{j+1}| \ri) \ri|    \nonumber\\
&&\leq \displaystyle \frac{c_{\eta}}{t^{3/2}}
\sum_{l=0}^{\infty} \alpha^l  \int_{\erre^6} dx_0 \, dx_{l+1} |f(x_0)| |g(x_{l+1})| \nonumber \\
&& \cdot \displaystyle \int_{\erre^{3l}} dx_1 \ldots dx_l \frac{ \prod_{j=1}^l | V_R(x_j )| }{ (2\pi)^{l+1}
\prod_{j=1}^l |x_j - x_{j+1}|} \sum_{j=0}^{l} |x_j - x_{j+1}|
\label{pappa}
\end{eqnarray}

\n
where $\eta$ is a cut-off function, i.e. a function $\eta:\erre^+ \rightarrow \erre$ such that $\eta\in C^{\infty}_0(\erre^{+})$, $\eta(x)=1$ for $0<x<1$, $\eta(x)=0$
for $x>2$. In (\ref{pappa}) we have used the estimate (lemma 2.4 in \cite{rs})

\begin{equation}
\sup_{L \geq 1}
\lf|\int_0^{\infty} d\la \,e^{it\la} \eta\lf(\frac{\sqrt{\la}}{L}\ri)
\sin \lf( \sqrt{\la} \sum_{j=0}^{l} |x_j - x_{j+1}| \ri) \ri| \leq  \frac{c_{\eta}}{t^{3/2}}
\sum_{j=0}^{l} |x_j - x_{j+1}|
\label{pseudodisp}
\end{equation}

\n
 where $c_{\eta}$ only depends on $\eta$. The last integral in (\ref{pappa}) can be estimated using the Kato norm of the potential (lemma 2.5 in \cite{rs}); moreover using (\ref{stimkato}) we have $(2 \pi)^{-1} \alpha \|V_R\|_{\mathcal{K}} <1$. Then

\ba
&&| <e^{-it \hat{h}(R)} f,g>|\leq  \frac{c_{\eta}}{2 \pi t^{3/2}} 
\sum_{l=0}^{\infty} \alpha^l  \int_{\erre^6} dx_0 \, dx_{l+1} |f(x_0)| |g(x_{l+1})| 
(l+1) \lf( \frac{  \| V_R \|_{\mathcal K} }{2\pi} \ri)^l 
\nonumber\\
&&
 \leq  \frac{C_0}{t^{3/2}} \| f\|_{L^1} \| g\|_{L^1}
\ea

\n
where

\be\label{C_0}
C_0 = \frac{c_{\eta}}{2 \pi} \sum_{l=0}^{\infty} (l+1) \left( \f{\alpha \|V\|_{\mathcal K}}{2 \pi}  \right)^{l}
\ee

\n
and this concludes the proof.
\end{dem}

\n
We shall now prove the uniform dispersive estimate for the derivatives of $e^{-i t \hat{h}(R)}$ with respect to the parameter $R$.

\n
For the proof of theorem 1, we only need derivatives up to order four but it is easy to extend the result to derivatives of any order.

\begin{prop}
Let us assume $V \in W^{\gamma,1} \cap H^\gamma$, $\gamma_1, 
\dots \gamma_n \in \natu$, $\sum_{i=1}^n \gamma_i = \gamma$, 
$m_1, \dots m_n \in \{ 1, \dots K \}$, 
$s_1, \dots s_n \in \{ 1, 2, 3 \}$, and let
\be\label{alfa*gamma}
\alpha^{*}_{\gamma} = \frac{ \pi^{2/3}}{3 \cdot 2^{\gamma -1}K} \,
 \|V\|^{-1/3}_{W^{\gamma,1}} \,
\|V\|^{- 2/3}_{H^{\gamma}}
\ee
Then for any $\alpha < \alpha^{*}_{\gamma}$ there 
exists a constant $C_{\gamma} > 0$ such that

\begin{equation}
\sup_{R} \|( \prod_{i=1}^n D^{\gamma_i}_{m_i,s_i} ) e^{-it \hat{h}(R)}  
\|_{ \mathcal L(L^1 , L^{\infty} )} \leq  \frac{C_{\gamma}}{t^{3/2}} 
\label{dispder}
\end{equation}
\end{prop}

\begin{dem}
The proof is a slight modification of the proof of proposition 3.1. 
In order to avoid a cumbersome notation, we limit the proof
to the case $n=2$. The general case can be proven in the same way.

\noindent
However, we stress that in the present paper we use inequality
(\ref{dispder}) in the cases $n=1$ and $n=2$ only.

\n
The first step is to show that the Born 
series of the resolvent can be differentiated term by term, i.e.

\ba
&&< D^{\gamma_1}_{m_1,s_1} D^{\gamma_2}_{m_2,s_2}
{\mathcal R}_R (\la + i0)f,g>  \; \;\nonumber\\ & & \nonumber \\
&&= \sum_{l=1}^{\infty} (\! -\alpha )^l \! 
\sum_{\gamma_{1,1} =0}^{\gamma_1} \sum_{\gamma_{1,2} =0}^{\gamma_{1,1}} \dots
\sum_{\gamma_{1,l-1} =0}^{\gamma_{1,l-2}}
\left( \!\!\!\begin{array}{c}
  \gamma_1 \\
\gamma_{1,1}
    \end{array} \!\!\! \right)
\left( \!\!\!\begin{array}{c}
  \gamma_{1,1} \\
\gamma_{1,2}
    \end{array} \!\!\! \right)
    \cdots
\left( \!\!\!\begin{array}{c}
  \gamma_{1,l-2} \\
\gamma_{1,l-1}
    \end{array} \!\!\! \right)
 \nonumber \\ & &
 \sum_{\gamma_{2,1} =0}^{\gamma_2} 
\sum_{\gamma_{2,2} =0}^{\gamma_{2,1}} \dots
\sum_{\gamma_{2,l-1} =0}^{\gamma_{2,l-2}}
\left( \!\!\!\begin{array}{c}
  \gamma_2 \\
\gamma_{2,1}
    \end{array} \!\!\! \right)
\left( \!\!\!\begin{array}{c}
  \gamma_{2,1} \\
\gamma_{2,2}
    \end{array} \!\!\! \right)
    \cdots
\left( \!\!\!\begin{array}{c}
  \gamma_{2,l-2} \\
\gamma_{2,l-1}
    \end{array} \!\!\! \right) 
\nonumber\\
& & \cdot <\! {\mathcal R}_0 (\la + i0) \! \lf( 
\! D^{\gamma_1 - \gamma_{1,1}}_{m_1,s_1} 
D^{\gamma_2 - \gamma_{2,1}}_{m_2,s_2}
 V_R \ri) 
\! {\mathcal R}_0 (\la + i0) \dots
\left( \! D^{\gamma_{1,l-1}}_{m_1,s_1} 
D^{\gamma_{2,l-1}}_{m_2,s_2} V_R \ri) 
\! {\mathcal R}_0 (\la + i0)
    f,g\!> \nonumber\\ 
& & \nonumber \\
&& =  \sum_{l=1}^{\infty} (\! -\alpha )^l \!\! \!\!
\sum_{{\substack{j_{1,1}\dots j_{1,l} \geq 0 \\ 
\sum_i j_{1,i} =\gamma_1}}} \  
\sum_{{\substack{j_{2,1}\dots j_{2,l} \geq 0 \\ 
\sum_i j_{2,i} =\gamma_2}}}
\!\! c_{j_{1,1} \ldots j_{1,l}} c_{j_{2,1} \ldots j_{2,l}}
\nonumber\\
& &\cdot <\! {\mathcal R}_0 (\la + i0) \!\lf( 
\! D^{j_{1,1}}_{m_1,s_1} 
D^{j_{2,1}}_{m_2,s_2}
V_R \ri) 
\! {\mathcal R}_0 (\la + i0) \cdots
 \lf( \! D^{j_{1,l}}_{m_1,s_1} 
D^{j_{2,l}}_{m_2,s_2}
V_R \ri) 
\! {\mathcal R}_0 (\la + i0) f,g\!> \nonumber\\
&&\label{derres}
\ea

\n
where
\begin{eqnarray}
&&c_{j_{k,1} \ldots j_{k,l}} = 
\left( \!\!\!\begin{array}{c}
  \sum_{i=1}^{l} j_{k,i} \\
  \sum_{i=2}^{l} j_{k,i}
    \end{array} \!\!\! \right)
  \left( \!\!\!\begin{array}{c}
  \sum_{i=2}^{l} j_{k,i} \\
  \sum_{i=3}^{l} j_{k,i}
    \end{array} \!\!\! \right)
\cdots
\left( \!\!\!\begin{array}{c}
 j_{k,l-1} + j_{k,l} \\
  j_{k,l}
    \end{array} \!\!\! \right),  ~~~~~
k=1,2
\label{coef}
\ea
and  the r.h.s. of \eqref{derres} is an absolutely convergent 
series and belongs to $L^1(\erre^+, d\la)$. In order to
prove this statement we estimate the $L^1$ 
norm of the general term of the series.
Denoting  $D^{j_{1,k}}_{m_1,s_1} D^{j_{2,k}}_{m_2,s_2} 
V_R$ by $V_R^{j_{1,k} j_{2,k}}$, we have
\ba
&&\int_0^{+\infty} d\la \, \lf|
<{\mathcal R}_0 (\la + i0)  V_R^{j_{1,1} j_{2,1}}
{\mathcal R}_0 (\la + i0) \cdots
V_R^{j_{1,l} j_{2,l}} {\mathcal R}_0 (\la + i0)f,g> \ri|
\nonumber\\
&&= \int_0^{+\infty} d\la \, \lf|< \lf[ \prod_{i=1}^{l-1} 
 {\rm{sgn}} \lf( V_R^{j_{1,i} j_{2,i}} \ri)
\lf| V_R^{j_{1,i} j_{2,i}}
 \ri|^{1/2}{\mathcal R}_0 (\la + i0)\left| V_R^{j_{1,i+1} 
j_{2,i+1}} \right|^{1/2} \ri] 
 \ri. \nonumber\\
&&\lf.  \cdot {\rm{sgn}} \left(V_R^{j_{1,l} j_{2,l}} \right)
\lf| V_R^{j_{1,l} j_{2,l}} \ri|^{1/2}
{\mathcal R}_0 (\la + i0) f, 
\lf| V_R^{j_{1,1} j_{2,1}} 
\ri|^{1/2}{\mathcal R}_0 (\la - i0)g> \ri|
\ea

\vs
\n
Using (\ref{HS}) and \eqref{rollstorto}, we have
\ba\label{prodK}
&&\lf\| \prod_{i=1}^{l-1} 
 {\rm{sgn}} \lf( V_R^{j_{1,i} j_{2,i}} \ri)
\lf| V_R^{j_{1,i} j_{2,i}}
 \ri|^{1/2}{\mathcal R}_0 (\la + i0)
\left| V_R^{j_{1,i+1} j_{2,i+1}} \right|^{1/2} 
\ri\|_{{\mathcal L} (L^2, L^2)}
\nonumber\\
&&\leq \left( \frac{c_1}{2 \pi}\right)^{l-1} \prod_{i=1}^{l-1} 
\left(  \|V_R^{j_{1,i} j_{2,i}} \|_{L^1}^{1/3} 
\|V_R^{j_{1,i} j_{2,i}} \|_{L^2}^{2/3}
\|V_R^{j_{1,i+1} j_{2,i+1}} \|_{L^1}^{1/3} 
\|V_R^{j_{1,i+1} j_{2,i+1}} \|_{L^2}^{2/3} 
\right)^{1/2} \nonumber\\
&&= \left( \frac{c_1}{2 \pi}\right)^{l-1}
\lf( \|V_R^{j_{1,1} j_{2,1}}\|_{L^1}^{1/3} 
\|V_R^{j_{1,1} j_{2,1}}\|_{L^2}^{2/3} \ri)^{1/2}
\lf( \|V_R^{j_{1,l} j_{2,l}}\|_{L^1}^{1/3} 
\|V_R^{j_{1,l} j_{2,l}}\|_{L^2}^{2/3} \ri)^{1/2} \nonumber \\
& & \prod_{i=2}^{l-1}\|V_R^{j_{1,i} j_{2,i}}\|_{L^1}^{1/3} 
\|V_R^{j_{1,i} j_{2,i}}\|_{L^2}^{2/3}
\ea

\n
Exploiting (\ref{prodK}), the 
  Schwartz's inequality, the Kato smoothing 
property (\ref{ksmooth}) and (\ref{stimroll}), we obtain
\ba\label{L1}
&&\int_0^{+\infty} d\la \, \lf|
<{\mathcal R}_0 (\la + i0) V_R^{j_{1,1} j_{2,1}}
{\mathcal R}_0 (\la + i0) \ldots
 V_R^{j_{1,l} j_{2,l}} {\mathcal R}_0 (\la + i0)f,g> \ri|
\nonumber\\
&&\leq
\left( \frac{c_1}{2 \pi}\right)^{l-1}
\lf( \|V_R^{j_{1,1} j_{2,1}}\|_{L^1}^{1/3} 
\|V_R^{j_{1,1} j_{2,1}}\|_{L^2}^{2/3} \ri)^{1/2}
\lf( \|V_R^{j_{1,l} j_{2,l}}\|_{L^1}^{1/3} 
\|V_R^{j_{1,l} j_{2,l}}\|_{L^2}^{2/3} \ri)^{1/2} \nonumber \\
& & \prod_{i=2}^{l-1}\|V_R^{j_{1,i}  j_{2,i}}\|_{L^1}^{1/3} 
\|V_R^{j_{1,i} j_{2,i}}\|_{L^2}^{2/3}
\left( \int_0^{+\infty} d\la \, \left\| 
\left| V_R^{j_{1,1} j_{2,1}}
\right|^{1/2} {\mathcal R}_0 ( \la - i 0 ) g 
\ri\|^{2}_{L^2} \ri)^{1/2} \nonumber \\
& &
\cdot \lf( \int_0^{+\infty} d\la \, \lf\| | V_R^{j_{1,l} j_{2,l}}|^{1/2} 
{\mathcal R}_0 ( \la + i 0 ) f \ri\|^{2}_{L^2} \ri)^{1/2}
\nonumber\\
&&\leq c  \frac{c_1^{l +1}}{(2 \pi)^{l-1}}  
\left( \prod_{i=1}^{l}
\|V_R^{j_{1,i} j_{2,i}}\|_{L^1}^{1/3} 
\|V_R^{j_{1,i} j_{2,i}}\|_{L^2}^{2/3} 
\right) \|f\|_{L^2} \|g\|_{L^2}\nonumber\\
&&\leq c \,  c_1   \left( \frac{K}{2 \pi}  c_1  
\|V\|_{W^{\gamma,1}}^{1/3} \|V\|_{H^{\gamma}}^{2/3} \right)^{l} 
\|f\|_{L^2} \|g\|_{L^2}
\ea
where in the last line we used 
$\|V^{j_{1,i} j_{2,i}}_{R} \|_{L^1} \leq K \|V\|_{W^{\gamma,1}}$ 
and $\|V^{j_{1,i} j_{2,i}}_{R} \|_{L^2} \leq K \|V\|_{H^{\gamma}}$.

\n
The  $L^1(\erre^+,d\la)$ norm of \eqref{derres} can 
be estimated as follows 
\ba
&&\int_0^{+\infty} d\la \, \lf|
<D^{\gamma_1}_{m_1,s_1}D^{\gamma_2}_{m_2,s_2}
{\mathcal R}_R (\la + i0)f,g> \ri| \nonumber\\
&& \leq c \, c_1   \|f\|_{L^2} \|g\|_{L^2} \sum_{l=1}^{\infty}
\lf( \alpha \frac{K}{2\pi}  c_1  
\|V\|_{W^{\gamma,1}}^{1/3} \|V\|_{H^{\gamma}}^{2/3}   \ri)^l
\sum_{{\substack{j_{1,1}\dots j_{1,l} \geq 0 \\ 
\sum_i j_{1,i} =\gamma_1}}} \  
\sum_{{\substack{j_{2,1}\dots j_{2,l} \geq 0 \\ 
\sum_i j_{2,i} =\gamma_2}}}
\!\! c_{j_{1,1} \dots j_{1,l}} c_{j_{2,1} \dots j_{2,l}}
\nonumber\\
&&\leq  c \, c_1   \|f\|_{L^2} \|g\|_{L^2} 
\sum_{l=1}^{\infty} l^{\gamma}
\lf( \alpha \frac{K}{2  \pi}  c_1  
\|V\|_{W^{\gamma,1}}^{1/3} \|V\|_{H^{\gamma}}^{2/3}   \ri)^l
\label{bddres}
\ea
where we used the identity 
\begin{equation}
\sum_{ \sum_i j_{k,i} =\gamma_k}
 c_{j_{k,1} \dots j_{k,l}} = l^{\gamma_k}, ~ ~ \ k = 1,2
\end{equation}
\n
Since the series in (\ref{bddres})  
converges for $\alpha < \alpha^{*}_{\gamma}$, 
we conclude that  (\ref{derres}) holds and 
the r.h.s.  is absolutely convergent and 
belongs to $L^{1}(\erre^+, d \lambda)$.

\noindent
Let us now consider the derivatives of the 
unitary group; using again the cut-off 
function $\eta$ as in \eqref{pappa},
we can write 
\begin{equation}
| <D^{\gamma_1}_{m_1,s_1} D^{\gamma_2}_{m_2,s_2} 
e^{-it \hat{h}(R)} f,g>| \leq
\sup_{L \geq 1} \lf|\int_0^{\infty} d\la \,e^{it\la} 
\, \eta\lf(\frac{\sqrt{\la}}{L}\ri) 
\Im <D^{\gamma_1}_{m_1,s_1} D^{\gamma_2}_{m_2,s_2} 
{\mathcal R}_R 
(\la + i0)f,g> \ri|
\label{pappa2}
\end{equation}

\n
Using (\ref{derres}), Fubini's theorem and  \eqref{pseudodisp}, we have 
\begin{eqnarray}
&&| <D^{\gamma_1}_{m_1,s_1} D^{\gamma_2}_{m_2,s_2} 
e^{-it \hat{h}(R)} f,g>|  
\leq \sum_{l=1}^{\infty} \alpha^l 
\sum_{{\substack{j_{1,1}\dots j_{1,l} \geq 0 \\ 
\sum_i j_{1,i} =\gamma_1}}} \  
\sum_{{\substack{j_{2,1}\dots j_{2,l} \geq 0 \\ 
\sum_i j_{2,i} =\gamma_2}}}
\!\! c_{j_{1,1} \ldots j_{1,l}} c_{j_{2,1} \ldots j_{2,l}}
\nonumber \\ & & 
\cdot \int_{\erre^6} dx_0 \, dx_{l+1} |f(x_0)| |g(x_{l+1})|
\nonumber \\ & & 
\cdot \int_{\erre^{3l}} \!\! dx_1 \ldots dx_l 
\frac{ \prod_{i=1}^l | V_R^{j_{1,i} j_{2,i}}(x_i )| }{ (2\pi)^{l+1}
\prod_{i=0}^l |x_i - x_{i+1}|}
\sup_{L \geq 1}
\lf|\int_0^{\infty} d\la \,e^{it\la} \,\eta\lf(\frac{\sqrt{\la}}{L}\ri)
\sin \lf( \sqrt{\la} \sum_{i=0}^{l} |x_i - x_{i+1}| \ri) \ri| \nonumber\\
&& \leq \frac{c_{\eta}}{t^{3/2}}  \sum_{l=1}^{\infty} \alpha^l  
\sum_{{\substack{j_{1,1}\dots j_{1,l} \geq 0 \\ 
\sum_i j_{1,i} =\gamma_1}}} \  
\sum_{{\substack{j_{2,1}\dots j_{2,l} \geq 0 \\ 
\sum_i j_{2,i} =\gamma_2}}}
\!\! c_{j_{1,1} \ldots j_{1,l}} c_{j_{2,1} \ldots j_{2,l}}
\int_{\erre^6} dx_0 \, dx_{l+1} 
|f(x_0)| |g(x_{l+1})| \nonumber\\
& & 	
\cdot \int_{\erre^{3l}} dx_1 \ldots dx_l
\frac{ \prod_{i=1}^l | V_R^{j_{1,i} 
j_{2,i}}(x_i )| }{ (2\pi)^{l+1} 
\prod_{i=0}^l |x_i - x_{i+1}|} 
\sum_{i=0}^{l} |x_i - x_{i+1}|
\end{eqnarray}

\vs
\n
Following the line of lemma 2.5 in \cite{rs}, 
the last integral can be dominated
   uniformly in $x_0$ and $x_{l+1}$
  using the Kato norm of the derivatives  
of the potential $V$.  In fact one obtains

\ba\label{stimDgamma}
&&| 
<D^{\gamma_1}_{m_1,s_1}D^{\gamma_2}_{m_2,s_2}
e^{-it \hat{h}(R)} f,g>| \nonumber \\ 
& & 
\leq  \frac{c_{\eta}}{2 \pi t^{3/2}} 
\|f\|_{L^1} \|g\|_{L^1} \sum_{l=1}^{\infty} 
\alpha^l  \frac{l +1}{(2 \pi)^l} 
\sum_{\substack{j_{1,1}\dots j_{1,l} \geq 0 \\
 \sum_i j_{1,i} =\gamma_1}}  \
\sum_{{\substack{j_{2,1}\dots j_{2,l} \geq 0 
\\ \sum_i j_{2,i} =\gamma_2}}}    
\!\! c_{j_{1,1} \ldots j_{1,l}} c_{j_{2,1} \ldots j_{2,l}}
\prod_{i=1}^l 
\|V^{j_{1,i} j_{2,i}}_R\|_{\mathcal{K}} \nonumber\\
&&\leq \frac{c_{\eta}}{2 \pi t^{3/2}} \|f\|_{L^1} 
\|g\|_{L^1} \sum_{l=1}^{\infty} (l+1) \left( 
\alpha \frac{K}{2 \pi} c_2 \|V\|^{1/3}_{W^{\gamma,1}} 
\|V\|^{2/3}_{H^{\gamma}} \right)^{l}
\sum_{{\substack{j_{1,1}\dots j_{1,l} \geq 0 \\ 
\sum_i j_{1,i} =\gamma_1}}} \  
\sum_{{\substack{j_{2,1}\dots j_{2,l} \geq 0 \\ 
\sum_i j_{2,i} =\gamma_2}}}
\!\! c_{j_{1,1} \ldots j_{1,l}} c_{j_{2,1} \ldots j_{2,l}}
\nonumber \\ 
& & 
\leq \frac{c_{\eta}}{2 \pi t^{3/2}} \|f\|_{L^1} 
\|g\|_{L^1} \sum_{l=0}^{\infty} (l+1) l^{\gamma} 
\left( \f{\alpha}{\alpha_{\gamma}^{*}}  \right)^{l}
\ea
where we used (\ref{stimkato}) and 
added the term $l=0$ in the last sum of (\ref{stimDgamma}). 

\n
Since $\alpha < \alpha^{*}_{\gamma}$, the 
series in (\ref{stimDgamma}) converges 
and we get (\ref{dispder}) with
\be\label{cgamma}
C_{\gamma} = \frac{c_{\eta}}{2 \pi }
\sum_{l=0}^{\infty} (l+1) l^{\gamma} \left( \f{
\alpha}{\alpha_{\gamma}^{*}} \right)^{l}
\ee
\end{dem}

\vs
\n
{\bf Remark}.  We observe  that $\alpha^{*}_{\gamma}$ is decreasing as a function of $\gamma$. Then for the proof of theorem 1 it is sufficient to choose $\alpha^{*} = \alpha^{*}_{4}$.

\n
We also notice that $C_\gamma$ is increasing as a function of $\gamma$, and this fact is  used during the proof of theorem 1.

\vs
\n
For the proof of theorem 1 we also need   a uniform $L^2$ estimate of the derivatives with respect to the parameter $R$ of the unitary group of the light particles.

\n
For a single light particle, exploiting the spectral theorem and (\ref{bddres}), we immediately get

\ba
&&\sup_{R} \left\| ( \prod_{i=1}^{n}  D^{\gamma_i}_{m_i,s_i})  e^{-it \hat{h}(R)} f \right\|_{L^2} \leq a_{\gamma}
\ea

\n
for any $\gamma$ integer (including zero), $\|f\|_{L^2} =1$ and $\alpha < \alpha^{*}$, where $a_{0} =1$ and 

\be
a_{\gamma}=   c \, c_1    \sum_{l=1}^{\infty} l^{\gamma}
\lf( \alpha \frac{K}{2  \pi}  c_1  \|V\|_{W^{\gamma,1}}^{1/3} \|V\|_{H^{\gamma}}^{2/3}   \ri)^l \;\;\;\;\;\;\; \gamma \neq 0
\ee

\n
For an arbitrary number $N$ of light particles, the Leibniz's rule yields

\be\label{L2prod}
\sup_{R}  \left\| ( \prod_{i=1}^{n} D^{\gamma_i}_{m_i,s_i} ) \prod_{k=1}^{N} e^{-it h_{k}(R)} \chi_k \right\|_{L^2(\erre^{3N})} \leq
\sum_{ \substack{j_1\ldots j_N =0 \\ \sum_i j_i =\gamma} }^{\gamma} \!\! c_{j_1 \ldots j_N}
a_{j_1} \cdots a_{j_N}
\ee

\vs
\n
where $\|\chi_k\|_{L^2} =1$. Notice that in the right hand side of (\ref{L2prod}) at most $\gamma$ of the constants $a_{j_i}$ are different from one and moreover each $a_{j_i}$ is less or equal to $a_{\gamma}$. Then we obtain the uniform $L^2$ estimate 

\be\label{uniL2}
\sup_{R}  \left\| ( \prod_{i=1}^{n} D^{\gamma_i}_{m_i,s_i} ) \prod_{k=1}^{N} e^{-it h_{k}(R)} \chi_k \right\|_{L^2(\erre^{3N})} \leq N^{\gamma} a_{\gamma}^{\gamma} \, \leq  \, 
\hat{C}_{\gamma}
\ee

\n
where we have defined

\be\label{hatC}
\hat{C}_{\gamma} = \max_{0 \leq \lambda \leq \gamma} N^{\lambda} a_{\lambda}^{\lambda}
\ee

\vs

\n
The proof of the uniform  estimates (\ref{disp}), (\ref{dispder}), (\ref{uniL2})  are based on a perturbative analysis and this requires the assumption of a small potential. We believe that this is only a technical limitation which could be removed with a more careful analysis. 

\n
In fact, following a different approach due to Yajima (\cite{y}),  uniform estimates  can be  easily proved in the simpler case $K=1$ for an arbitrarily large potential.

\n
In this case the crucial ingredient is  the boundedness of  the wave operators 
in  the Sobolev spaces $W^{k,p}$.

\begin{prop}
Let $K=1$, $\gamma \in \natu$, $p \in [2, \infty]$  and let us assume that:

\n
i) $V \in W^{\gamma, \infty}_{\delta}$,  
 for    $\delta >5$;

\n
ii)   $V\geq 0$; 

\n
iii) $g \in L^2 \cap W^{\gamma,q}$, $\gamma \in \natu$, $q^{-1} = 1 - p^{-1}$.

\n
 Then there exists a constant $b_{\gamma,p,q}>0$ such that
 
\begin{equation}
\sup_R \| D^{\gamma}_s e^{-it \hat{h}(R)} g \|_{L^{p} } \leq \frac{b_{\gamma,p,q}}{t^{3(2-q)/2q}} \| g\|_{W^{\gamma,q}}
\label{yadisp}
\end{equation}
\end{prop}
\begin{dem}
Since $K=1$,  the dependence of $e^{-it \hat{h}(R)}$ on the parameter $R \in \erre^3$ can be extracted using the unitary translation operator $T_R$, $(T_R f)(x)=f(x+R) $. 
Moreover, using the intertwining property of the wave operators one has

\begin{equation}
e^{-it\hat{ h}(R)} = T_R \,\Omega_+ \,e^{-it h_0} \,\Omega_+^{-1} \,T^{-1}_{R}
\label{trasl}
\end{equation}

\n
where $\Omega_{+}$ is the wave operator for the pair $( \hat{h}(0) , h_0)$.

\n
We use \eqref{trasl} to compute the derivatives with respect to the parameter  $R$, noticing that 

\begin{equation}
D_{s} \, T_R = - T_R \,d_s \, , \qquad D_{s} \, T^{-1}_R = T^{-1}_R\, d_s
\end{equation}

\n
where $(d_s^i  f)(x)= \frac{\partial^i f}{\partial x_s^i}(x)$ and $x_s$ is the $s$-th component of $x$.

\n
We have 

\begin{equation}
D^{\gamma}_{s} \,e^{-it \hat{h}(R)} g = \sum_{k=0}^{\gamma} (-1)^{k} \binom{\gamma}{k} T_R \,d^k_s\,
\Omega_+ \,e^{-it h_0} \,\Omega_+^{-1} \,d^{\gamma-k}_s \,T_R^{-1}g
\label{omega}
\end{equation}

\vs
\n
The generic  term of the sum in \eqref{omega} can be estimated as follows

\ba
&&
\lf\| T_R \,d^k_s\,\Omega_+ \,e^{-it h_0} \,\Omega_+^{-1} \,d^{\gamma-k}_s \,T_R^{-1}g \ri\|_{L^{p}} =
\lf\| d^k_s\,\Omega_+ \,e^{-it h_0} \,\Omega_+^{-1} \,d^{\gamma -k}_s \,T_R^{-1}g \ri\|_{L^{p}} \nonumber\\
&&\leq c_{k,p}(\Omega_+) \lf\| e^{-it h_0} \,\Omega_+^{-1} \,d^{\gamma -k}_s \,T_R^{-1}g \ri\|_{W^{k,p}} 
\leq
 \frac{c}{t^{3(2-q)/2q} }   c_{k,p}(\Omega_+)  \lf\|\Omega_+^{-1}\, d^{\gamma - k}_s \,T_R^{-1}g \ri\|_{W^{k,q}}
\nonumber\\
&&\leq
 \frac{c}{t^{3(2-q)/2q} } c_{k,p}(\Omega_+) c_{k,q}(\Omega_+^{-1})   \lf\| d^{\gamma - k}_s \,T_R^{-1}g \ri\|_{W^{k,q}}
\leq \frac{c}{t^{3(2-q)/2q} } c_{k,p}(\Omega_+) c_{k,q}(\Omega_+^{-1})   \lf\| T_R^{-1}g \ri\|_{W^{\gamma,q}}\nonumber\\
&&=\frac{c}{t^{3(2-q)/2q} } c_{k,p}(\Omega_+) c_{k,q}(\Omega_+^{-1})   \lf\| g \ri\|_{W^{\gamma,q}}\label{omega1}
\ea

\vs
\n
where we have used the isometric character of $T_R$, the boundedness of the wave operators in $W^{k,p}$ (see \cite{y}),  the fact that the free propagator commutes with derivatives and the standard 
estimate for the free Schr\"{o}dinger group

\be
\left\| e^{-it h_0} \right\|_{\mathcal L (L^q , L^p)} \leq \frac{c}{t^{3(2-q)/2q}}
\ee

\vs
\n
 In (\ref{omega1}) the symbols $c_{k,p}(\Omega_{+})$, $c_{k,p}(\Omega_{+}^{-1})$ denote the operator norm in $W^{k,p}$ of $\Omega_{+}$, $\Omega_{+}^{-1}$ respectively. 
From \eqref{omega} and (\ref{omega1}) we obtain the proof of \eqref{yadisp}
\end{dem}

\vs
\n
{\bf Remark.}  Notice that for $p=\infty$ the estimate (\ref{yadisp}) reduces to the uniform dispersive estimate and, in this case, we denote

\be\label{bgamma}
B_\gamma = \max_{0 \leq \lambda \leq \gamma} b_{\lambda, \infty,1}
\ee

\vs
\n
Moreover for $p=2$, proceeding as in (\ref{L2prod}), (\ref{uniL2}), we easily get

\be\label{bhatgamma}
\sup_R \left\| D^{\gamma}_{s} \prod_{k=1}^{N} e^{-it h_k(R)} \chi_k \right\|_{L^2(\erre^{3N})} \leq N^{\gamma} b^{\gamma}_{\gamma,2,2} \left( \max_j  \|\chi_j \|_{H^{\gamma}} \right)^{\gamma}  \leq \hat{B}_{\gamma} 
\ee

\n
where we have defined 

\be\label{bhatgamma1}
\hat{B}_{\gamma} = \max_{0 \leq \lambda \leq \gamma}
N^{\lambda} b^{\lambda}_{\lambda,2,2} \left( \max_j  \|\chi_j \|_{H^{\lambda}} \right)^{\lambda}
\ee


\vs
\vs
\vs

\section{Some commutator estimates  involving 
the unitary group $e^{-itX}$}

\vs

\noindent
In this section we discuss some estimates for  the 
commutator of the unitary group $e^{-itX}$ 
with the operators $X_0$ and $R_l^2$ in the Hilbert space of the heavy particles.

\n
Such estimates  are repeatedly used in the proof of theorem 1.  Since we have not found them in the literature,  a simple proof is exhibited here for the convenience of the reader.

\n
We find convenient 
to express the results in terms of the weighted Sobolev space related to the $l$-th heavy particle, which were defined in the introduction.

\n
The first result concerns the commutator $[X_{0,l}, e^{-itX}]$.

\begin{prop}
Given $l \in \{1, \dots, K \}$, $f \in 
H^1_{l,0}(\erre^{3K})$ and $T>0$,  there exists a constant $\tilde{C} >0$ such that  
\begin{equation}
\label{komm1}
\| [ X_{0,l}, e^{-itX}] f \|_{L^2 (\erre^{3K})} \ 
\leq t \; \tilde C \| f \|_{H^1_{l,0}(\erre^{3K})}
\end{equation} 
for  any $t \in [0,T]$.
\end{prop}

\begin{dem}
For $s=1,2,3$ and using  the short-hand notation 
$\eta(t) = e^{-itX} f$,  a direct computation gives
\begin{equation}
i \frac {\partial } {\partial t} D_{l,s} \eta (t)
= X D_{l,s} \eta (t) + (D_{l,s} U) 
\eta (t)
\end{equation}
Therefore, by Duhamel's formula
\begin{equation}
D_{l,s} \eta (t) =  e^{-itX}D_{l,s} f - i \int_0^t d\tau e^{-i 
(t-\tau) X}
(D_{l,s} U) \eta(\tau)
\label{primader}
\end{equation}
Iterating the procedure one finds

\begin{eqnarray}
&&D_{l,s}^2 \eta (t)  =    e^{-itX}D_{l,s}^2 f - 
i \int_0^t d\tau e^{-i (t-\tau) X} (D_{l,s}^2 U) \eta(\tau)
 - 2 i  \int_0^t d\tau e^{-i (t-\tau) X} (D_{l,s} U) e^{-i \tau X} 
D_{l,s} f \nonumber \\
& & - 2  \int_0^t d\tau \int_0^\tau d\sigma  e^{-i (t-\tau) X} 
(D_{l,s} U)  e^{-i (\tau - \sigma) X} (D_{l,s} U) \eta(\sigma)
\end{eqnarray}
Therefore
\begin{eqnarray}
&&X_{0,l} \eta (t)  =    e^{-itX} X_{0,l} f - 
i \!\!\int_0^t \! d\tau e^{-i (t-\tau) X} (X_{0,l} U) \eta(\tau)
 - 2 i \sum_{s=1}^3 \int_0^t \! d\tau e^{-i (t-\tau) X} (D_{l,s} U) 
e^{-i \tau X} 
D_{l,s} f \nonumber \\
& & - 2  \sum_{s=1}^3 \int_0^t d\tau \int_0^\tau d\sigma  e^{-i 
(t-\tau) X} 
(D_{l,s} U)  e^{-i (\tau - \sigma) X} (D_{l,s} U) \eta(\sigma)
\end{eqnarray}

\n
Recalling the definition of $\eta(t)$, it follows 

\begin{eqnarray}
& & \| [ X_{0,l}, e^{-itX}] f \|_{L^2 (\erre^{3K})} \nonumber\\
&& \leq  \int_0^t d\tau \| (X_{0,l} U)  e^{-i \tau X} f   
\|_{L^2 (\erre^{3K})}  
+ 2 \sum_{s=1}^3 \int_0^t d\tau 
\| D_{l,s} U \|_{L^\infty (\erre^{3K})} \| D_{l,s} f \|_{L^2 
(\erre^{3K})}
\nonumber \\
& & + 2   \sum_{s=1}^3 \int_0^t d\tau  \int_0^\tau d\sigma 
\| D_{l,s} U \|_{L^\infty (\erre^{3K})}^2 \|   e^{-i \sigma X} f   
\|_{L^2 (\erre^{3K})} \nonumber \\
& &\leq  t \| \Delta U_l \|_{L^\infty} \| f \|_{L^2 (\erre^{3K})} +
2t \| U_l \|_{W^{1, \infty}}  \| f \|_{H^{1}_{l,0} (\erre^{3K})}
+ t^2  \| U_l \|_{W^{1, \infty}}^2 
\| f \|_{L^2 (\erre^{3K})}
\nonumber \\
& &\leq  t\,  \tilde C  \| f \|_{H^{1}_{l,0} (\erre^{3K})}
\end{eqnarray}

\vs
\n
where we used the fact that $D_{l,s} U = D_s U_l$ and
defined

\begin{equation}\label{Ctilde}
 \tilde C =  \max_{l} \left(\| \Delta U_l \|_{L^\infty} + 2  \| U_l \|_{W^{1, 
\infty}} 
+  T  \| U_l \|_{W^{1, \infty}}^2 \right)
\end{equation}
\end{dem}

\vs
\begin{cor} \label{quasiyajima}
For any $t \geq 0$ the operator
 $e^{-itX}$ is continuous in $W_{l,0}^{m,p} (\erre^{3K})$, with $l 
\in \{1, \dots, K\}$,
$m \in \natu$, $p \geq 1$.
\end{cor}
\begin{dem}
From (\ref{primader}) the following estimate is easily obtained
\begin{equation}
\| D_{l,s} e^{-itX} f \|_{L^p(\erre^{3K})} \leq \| D_{l,s} f 
\|_{L^p(\erre^{3K})}
+ t \| D_{s} U_l \|_{L^\infty}
\| f \|_{L^p(\erre^{3K})}
\end{equation}
and continuity in $W^{1,p}_{l,0}$ immediately follows.
For the case $m > 1$ the result is achieved differentiating
the quantity $e^{-itX} f$. 
\end{dem}

\vs
\n
The second result in this section is an estimate of the commutator $[R^2_l , e^{-itX}]$.

\begin{prop}\label{prop1}
 Given $l \in \{1, \dots K\}$, $f \in H^2_{l,2} 
(\erre^{3K})$ and $T>0$, there exists a  constant $\bar{C}_1 >0$ such that 

\begin{equation}
\| [ R_l^2, e^{-itX} ] f \|_{L^2 (\erre^{3K})} \ \leq \
t \, \bar{C_1} \|  f \|_{H^{2}_{l,2} (\erre^{3K})}
\end{equation}
for any $t \in [0,T]$.
\end{prop}

\begin{dem}
First, we observe that
\begin{equation}
 [ R_l^2, e^{-itX_0} ] = t  e^{-itX_0} S_{0,l} (t) \label{martello}
\end{equation}
where
\begin{equation}
S_{0,l} (t) = 
\label{alabarda1}
 - \sum_{s=1}^3 \left( 2i R_{l,s} D_{l,s} + t D^2_{l,s} \right) -3i
\end{equation}

\n
Formula (\ref{martello}) can be easily derived integrating by parts in the explicit integral representation of the free unitary  group. The action of the operator $S_{0,l}(t)$ is estimated as follows

\begin{eqnarray}
&&\| S_{0,l} (t) f \|_{L^2(\erre^{3K})}  \leq  2 \sum_{s=1}^3 
\| R_{l,s} D_{l,s} f \|_{L^2(\erre^{3K})}
 + t \sum_{s=1}^3 \| D^2_{l,s} f \|_{L^2(\erre^{3K})} + 3
\| f \|_{L^2(\erre^{3K})} \nonumber \\
&& \leq  2 \|  f \|_{H^{1}_{l,1}(\erre^{3K})} + t  \| f 
\|_{H^{2}_{l,0}(\erre^{3K})} + 3 \| f \|_{L^{2}(\erre^{3K})}
 \leq  c \,  (1+t) \|  f \|_{H^{2}_{l,1}(\erre^{3K})} \label{sol}
\end{eqnarray}

\vs
\n
Using Duhamel's formula,  the action of $[R_l^2, e^{-itX}]$ reads 

\begin{eqnarray}
&&[R_l^2, e^{-itX}]f  =  t  e^{-itX_0} S_{0,l} (t) f -i
\int_0^t d\sigma \, \sigma e^{-i\sigma X_0}  S_{0,l} (\sigma)U
 e^{-i(t-\sigma)X} f \nonumber \\
& & -i 
\int_0^t d\sigma \, e^{-i(t-\sigma)X_0} R_l^2 U e^{-i\sigma X} f 
+ i \int_0^t d\sigma \, e^{-i(t-\sigma)X_0} U e^{-i\sigma X} 
 R_l^2 f  \label{duham3}
\end{eqnarray}

\n
Estimate (\ref{sol}) directly applies to the first 
term of (\ref{duham3}), while for the second it gives

\begin{eqnarray}
\left\|  \int_0^t d\sigma \, \sigma e^{-i\sigma X_0}  S_{0,l} 
(\sigma)U
 e^{-i(t-\sigma)X} f \right\|_{L^2(\erre^{3K})} & \leq &
c  \int_0^t d\sigma \, \sigma (1 + \sigma) \|  U e^{-i(t-\sigma)X}f 
\|_{H^{2}_{l,1}(\erre^{3K})}
\nonumber \\ \label{erst} \end{eqnarray}
Notice that
\begin{eqnarray}
& & \|  U e^{-i(t-\sigma)X}f \|_{H^{2}_{l,1}(\erre^{3K})} \nonumber \\
&& =
 \sum_{\gamma_1 =0}^2 \sum_{\gamma_2 =0}^{2 - \gamma_1}
 \sum_{\gamma_3 =0}^{2 - \gamma_1 - \gamma_2} \| <R_l> 
D_{l,1}^{\gamma_1}
D_{l,2}^{\gamma_2} D_{l,3}^{\gamma_3}  \, U \, e^{-i(t-\sigma)X} f 
\|_{L^{2}(\erre^{3K})} \nonumber \\
& &\leq  8  \sum_{\gamma_1 =0}^{2}  \sum_{\gamma_2 =0}^{2 - \gamma_1}
 \sum_{\gamma_3 =0}^{2 - \gamma_1 - \gamma_2}   \sum_{\lambda_1 = 
0}^{\gamma_1}
 \sum_{\lambda_2 = 0}^{\gamma_2}  \sum_{\lambda_3 = 0}^{\gamma_3}
 \| <R_l> D_{l,1}^{\lambda_1} D_{l,2}^{\lambda_2} D_{l,3}^{\lambda_3}
U \|_{L^{\infty}(\erre^{3K})}  \nonumber \\
& & 
 \cdot \|  D_{l,1}^{\gamma_1 - \lambda_1}   D_{l,2}^{\gamma_2 - 
\lambda_2}   
D_{l,3}^{\gamma_3 - \lambda_3} 
e^{-i(t-\sigma)X} f \|_{L^{2}(\erre^{3K})} 
 \nonumber \\
& &\leq  c \, \| U \|_{W^{2, \infty}_{l,1} (\erre^{3K})}  \| 
e^{-i(t-\sigma)X} f  
\|_{H^{2}_{l,0}(\erre^{3K})}  \nonumber \\
&& \leq  c \, \| U \|_{W^{2, \infty}_{l,1} (\erre^{3K})}  \| f  
\|_{H^{2}_{l,0}(\erre^{3K})} \label{stimella}
\end{eqnarray}
where, in the last step we used Corollary \ref{quasiyajima}. 

\noindent
Thus, going back to (\ref{erst})
\begin{eqnarray}
\left\|  \int_0^t  \! d\sigma \, \sigma e^{-i\sigma X_0}  S_{0,l} 
(\sigma)U
 e^{-i(t-\sigma)X} f \right\|_{L^2(\erre^{3K})} \!\! \!& \leq & \!\! \!
c \left( \f {t^2} 2 + \f  {t^3} 3 \right) \!
 \| U \|_{W^{2, \infty}_{l,1} (\erre^{3K})} \| f  
\|_{H^{2}_{l,0}(\erre^{3K})}  \label{zweite}
\end{eqnarray}
The third term in (\ref{duham3}) can be estimated as follows
\begin{eqnarray}
\left\|  \int_0^t d\sigma \,  e^{-i(t-\sigma)X_0} R_l^2 U  
e^{-i\sigma X}
f \right\|_{L^2(\erre^{3K})} & \leq & 
t \, \|  R_l^2 U \|_{L^{\infty}(\erre^{3K})} \| f \|_{L^{2}(\erre^{3K})} 
\label{dritte}
\end{eqnarray}
Finally, the fourth term in  (\ref{duham3}) gives
\begin{eqnarray}
\left\|  \int_0^t d\sigma \,  e^{-i(t-\sigma)X_0} U  e^{-i\sigma X}
R_l^2 f \right\|_{L^{2}(\erre^{3K})}
& \leq & t \,
 \|  U \|_{L^{\infty}(\erre^{3K})} \| f \|_{H^{0}_{l,2}(\erre^{3K})} 
\label{vierte}
\end{eqnarray}
By (\ref{sol}) , (\ref{zweite}), (\ref{dritte}), (\ref{vierte}) we 
 conclude
\begin{eqnarray}
\left\| [ R_l^2,  e^{-itX}] f \right\|_{L^{2}(\erre^{3K})} 
& \leq & t \,\bar{C_1}  \| f \|_{H^{2}_{l,2}(\erre^{3K})} 
\end{eqnarray}
where we defined
\begin{equation}
 \bar{C_1} \ = \ c \max_{l} \left( 1 + T + T^2 \| U \|_{W^{2, \infty}_{l,1} 
(\erre^{3K})}
+  \| R_l^2 U \|_{L^{\infty} (\erre^{3K})} +  \| U 
\|_{L^{\infty}(\erre^{3K})}
\right)
\end{equation}
\end{dem}

\n
The last estimate  concerns the same commutator of the previous proposition, composed with the Laplacian with respect to $R_l$.
\begin{prop}
 Given $l \in \{1, \dots K\}$, $f \in H^4_{l,2} 
(\erre^{3K})$ and $T>0$ there exists a constant $\bar{C}>0$ such that
\begin{equation}
\|( X_{0,l} +I)  [ R_l^2, e^{-itX} ] f \|_{L^2 (\erre^{3K})} 
\ \leq \
t \,\bar{C} \|  f \|_{H^{4}_{l,2}(\erre^{3K})} 
\end{equation}
for any $t \in [0,T]$.
\end{prop}
\begin{dem}
From (\ref{duham3}) one has
\begin{eqnarray}
&&X_{0,l} [R_l^2, e^{-itX}]f  =  t  e^{-itX_0} X_{0,l} S_{0,l} (t) f 
-i
\int_0^t d\sigma \, \sigma e^{-i\sigma X_0}  X_{0,l} S_{0,l} (\sigma)U
 e^{-i(t-\sigma)X} f \nonumber \\
& & -i 
\int_0^t d\sigma \, e^{-i(t-\sigma)X_0} X_{0,l} R_l^2 U e^{-i\sigma 
X} f 
+ i \int_0^t d\sigma \, e^{-i(t-\sigma)X_0}  X_{0,l} U e^{-i\sigma X} 
 R_l^2 f  \nonumber \\ \label{duham4}
\end{eqnarray}
We estimate the first term in (\ref{duham4}) as follows
\begin{eqnarray} & &
\| t  e^{-itX_0} X_{0,l} S_{0,l} (t) f \|_{L^2 (\erre^{3K})}
 \nonumber \\
& &\leq  t \left\| \sum_{s'=1}^3 D^2_{l,s'} \left[ \sum_{s=0}^3 (2 i 
R_{l,s}
D_{l,s} f + t D^2_{l,s} f) - 3 i f \right]  \right\|_{L^2 
(\erre^{3K})}
 \nonumber \\
& &\leq  2t \sum_{s'=1}^3 \sum_{s=0}^3 \| D^2_{l,s'} R_{l,s} D_{l,s} 
f  
\|_{L^2 (\erre^{3K})} + t^2 \sum_{s'=1}^3 \sum_{s=0}^3 \| D^2_{l,s'}
D_{l,s}^2 f \|_{L^2 (\erre^{3K})}  + 3t \sum_{s'=1}^3 \| D^2_{l,s'}f \|_{L^2 (\erre^{3K})} \nonumber 
\\
& &\leq  4t \sum_{s'=1}^3 \sum_{s=0}^3 \| \delta_{s,s'} 
D_{l,s'}D_{l,s} f  
\|_{L^2 (\erre^{3K})} + 2t \sum_{s'=1}^3 \sum_{s=0}^3 \| 
R_{l,s}    D^2_{l,s'}
D_{l,s} f \|_{L^2 (\erre^{3K})} \nonumber \\
& & + t^2 \| f \|_{H^4_{l,0}(\erre^{3K})} + 3t \| f 
\|_{H^2_{l,0}(\erre^{3K})}
 \nonumber \\
&& \leq  7t \sum_{s=0}^3 \| f  
\|_{H^2_{l,0} (\erre^{3K})} + 2t  \| f  \|_{H^3_{l,1} (\erre^{3K})}
 + t^2 \| f \|_{H^4_{l,0}(\erre^{3K})} 
 \nonumber \\
& &\leq  c \, t (1+t) \| f \|_{H^4_{l,1}(\erre^{3K})}
\label{primeiro}
\end{eqnarray}
To estimate the second term in (\ref{duham4}) we exploit 
(\ref{primeiro}) and then proceed as in (\ref{stimella})
obtaining
\begin{eqnarray}
&&\left\|
\int_0^t d\sigma \, \sigma e^{-i\sigma X_0}  X_{0,l} S_{0,l} (\sigma) 
U
 e^{-i(t-\sigma)X} f \right\|_{L^2 (\erre^{3K})}
 \leq  c \int_0^t d\sigma \, \sigma (1+\sigma) \| U 
e^{-i(t-\sigma)X} f \|
_{H^4_{l,1} (\erre^{3K})}  \nonumber \\ 
&&\leq c
\int_0^t d\sigma \sigma (1 + \sigma) \, \| U \|_{W^{4, 
\infty}_{l,1}(\erre^{3K})} 
\|  e^{-i(t-\sigma)X} f \|
_{H^4_{l,0} (\erre^{3K})}
\leq  c t (t+t^2) \| U \|_{W^{4, \infty}_{l,1}(\erre^{3K})} \|f \|
_{H^4_{l,0} (\erre^{3K})} \nonumber \\ \label{segundo}
\end{eqnarray}
For the third term in (\ref{duham4}) we have
\begin{eqnarray}
& & \left\|  \int_0^t d\sigma e^{-i(t-\sigma)X_0} X_{0,l}
R_l^2 U e^{-i \sigma X} f \right\|_{L^2 (\erre^{3K})} \nonumber \\
& &\leq  \int_0^t d\sigma \left[ \|  (X_{0,l}
R_l^2 U) e^{-i \sigma X} f\|_{L^2 (\erre^{3K})} + 2 \sum_{s=1}^3
 \| D_{l,s} (R_l^2 U) D_{l,s}e^{-i \sigma X} f \|
_{L^2 (\erre^{3K})} \nonumber \right. \\ 
& &  + \| R_l^2 U X_{0,l} e^{-i \sigma X} f \|
_{L^2 (\erre^{3K})} \Bigg] \nonumber \\
& &\leq  t  \, \|  X_{0,l}
R_l^2 U \|_{L^\infty (\erre^{3K})} \| f \|_{L^2 (\erre^{3K})} +
 2 \sum_{s=1}^3   \| D_{l,s} (R_l^2 U)  \|_{L^\infty (\erre^{3K})} 
\int_0^t d\sigma \, 
 \|  D_{l,s}e^{-i \sigma X} f \|
_{L^2 (\erre^{3K})}  \nonumber \\ 
& &  +   \| R_l^2 U \|_{L^\infty (\erre^{3K})} \int_0^t d\sigma \,
 \|  X_{0,l} e^{-i \sigma X} f \|
_{L^2 (\erre^{3K})} \nonumber \\
& &\leq  c\,  t  \left( \| U_l \|_{W^{2,\infty}_{l,2}}  \| f \|_{L^2 
(\erre^{3K})} +
  \| U_l \|_{W^{1,\infty}_{l,2}} \| f \|_{H^1_{l,0}(\erre^{3K})} + 
 \| U \|
_{W^{0,\infty}_{l,2} (\erre^{3K})}  \| f \|_{H^2_{l,0}(\erre^{3K})} \right)
 \nonumber \\
& &\leq  c\,   t \, \| U \|
_{W^{2,\infty}_{l,2} (\erre^{3K})}  \| f \|_{H^2_{l,0} (\erre^{3K})}
\label{terco}
\end{eqnarray}

\n
For  the fourth term in (\ref{duham4}) we 
have
\begin{eqnarray}
& & \left\|  \int_0^t d\sigma e^{-i (t - \sigma) X_0}
X_{0,l} U e^{-i \sigma X} R^2_l f \right\|_{L^2 (\erre^{3K})}
\nonumber \\
& &\leq   \int_0^t d\sigma \left[ \, \|  ( X_{0,l}
U ) e^{-i \sigma X} R_l^2 f\|_{L^2 (\erre^{3K})} + 2 \sum_{s=1}^3
 \| (D_{l,s} U) D_{l,s}e^{-i \sigma X} R_l^2 f \|
_{L^2 (\erre^{3K})} \nonumber \right. \\ 
& & + \| U X_{0,l} e^{-i \sigma X}  R_l^2 f \|
_{L^2 (\erre^{3K})} \Bigg] \nonumber 
\ea
\ba
& &\leq  t \, \|  X_{0,l}
U \|_{L^\infty (\erre^{3K})} \|  R_l^2 f \|_{L^2 (\erre^{3K})} +
 2 \sum_{s=1}^3   \| D_{l,s}  U  \|_{L^\infty (\erre^{3K})} \int_0^t 
d\sigma \, 
 \|  D_{l,s}e^{-i \sigma X} R_l^2 f \|
_{L^2 (\erre^{3K})}  \nonumber \\ 
& & 
 +   \|  U \|_{L^\infty (\erre^{3K})} \int_0^t d\sigma \,
 \|  X_{0,l} e^{-i \sigma X} R_l^2 f \|
_{L^2 (\erre^{3K})} \nonumber \\
&& \leq  c\, t\, \left(   \| U_l \|_{W^{2,\infty}}  \| f \|_{H^{0,2}_{l,2} 
(\erre^{3K})} +
  \| U_l \|_{W^{1,\infty}} \| f \|_{H^1_{l,2}(\erre^{3K})} +   
\| U \|
_{L^{\infty} (\erre^{3K})}  \| f \|_{H^2_{l,2}(\erre^{3K})} \right) 
 \nonumber \\
& &\leq  c \, t \| U \|
_{W^{2,\infty}_{l,0} (\erre^{3K})}  \| f \|_{H^2_{l,2}(\erre^{3K})}
\label{cuarto}
\end{eqnarray}

\vs
\n
Therefore, by (\ref{duham4}), (\ref{primeiro}), (\ref{segundo}), 
(\ref{terco}),
(\ref{cuarto}) and proposition \ref{prop1} we finally obtain
\begin{eqnarray}
\| (X_{0,l} +I) [R_l^2, e^{-itX}] f \|_{L^2 (\erre^{3K})}
 & \leq & t \,\bar{C} \| f \|_{H^{4,2}_{l,2} (\erre^{3K})}
\end{eqnarray}
where
\begin{equation}\label{Cbar}
\bar{C} = \bar{C}_1 +  c \, \max_{l} \left[(1+T)\left( 1 + T \, \| U \|_{W^{4, \infty}_{l,2}  
(\erre^{3K})} \right)
+ \| U \|_{W^{2, \infty}_{l,2} (\erre^{3K})} +  \| U \|_{W^{2, 
\infty}_{l,0}(\erre^{3K})}\right]
\end{equation}
\end{dem}


\vs
\vs
\vs

\section{Application to decoherence}
\vs
\n
Some of the most peculiar
aspects of
 Quantum Mechanics are direct consequences of the superposition principle, 
i.e. the fact   that the normalized superposition of two quantum states is
a possible state for a
  quantum system. Interference effects between the two states and their consequences
on   the statistics of the expected results of a measurement performed on the
system do not have
  any explanation within the realm of classical probability theory.

  \n
 On the other hand this highly non-classical behaviour is extremely sensitive
to the
 interaction with the environment. The mechanism of irreversible diffusion
of quantum
  correlations in the environment is generally referred to as decoherence.
The analysis
  of this phenomenon within the frame of Quantum Theory is of great interest
and, at the  same time, of great difficulty inasmuch as results about the
dynamics of  large 
quantum systems are required in order to build up non-trivial models of
environment.

 \n
In this section we consider the mechanism of decoherence on a heavy particles
(the
  system) scattered by $N$ light particles (the environment). For this purpose
we follow  closely the line of reasoning of Joos and Zeh (\cite{jz}) and
we exploit formula
  (\ref{psia}) for the asymptotic wave function in the simpler case $U=0$.
  
  \n
  (For other rigorous analysis of the mechanism of decoherence see e.g. \cite{d}, \cite{ds}, \cite{ccf}).

\n
All the information concerning the dynamical behaviour of observables associated
with
the heavy particle is contained in the reduced density matrix, which in
our case is the
 positive,  trace class operator
 $\rho^{\ve}(t)$ in $L^{2}(\erre^{3})$ with ${\rm Tr} \,  \rho^{\ve}(t)=1$
 and integral kernel given by

\be \rho^{\ve}(t;R,R') = \int_{\erre^{3N}} dr \, \Psi^{\ve}(t;R,r)
\overline{\Psi^{\ve}}(t;R',r) \ee

\n An immediate  consequence of theorems 1, 1$'$ is that  for $\ve \rightarrow 0$ the operator $\rho^{\ve}(t)$ converges in the trace class norm to the asymptotic reduced density matrix

\be \rho^{a}(t) = e^{-it X_0} \rho_{0}^{a} e^{it X_0 }\\ \ee

\medskip

\n 
 where $\rho_{0}^{a}$ is a density matrix whose
integral kernel is

\ba &&\rho_{0}^{a}(R,R')= \phi(R) \overline{\phi}(R') {\mathcal
I}(R,R')\\
& & \nonumber \\
&&{\mathcal I}(R,R')= \prod_{j=1}^{N} \left( \Omega_{+}(R')^{-1} \chi_j,
\Omega_{+}(R)^{-1}  \chi_j \right)_{L^2}
\ea

\n
and $(\cdot, \cdot)_{L^2}$ denotes the scalar product in $L^2(\erre^3)$.

\n
Notice that the asymptotic dynamics of the heavy particle described by
 $\rho^{a}(t)$ is generated by $X_0$, i.e. the Hamiltonian of the heavy particle
when the light particles are absent. The effect of the interaction with
the light particles is expressed in the change of the initial state from
  $\phi(R)\overline{\phi}(R')$ to
 $\phi(R)\overline{\phi}(R') {\mathcal I} (R,R')$. Significantly the new
initial state is  not in product form, meaning that entanglement between
the system and the environment has taken place. Yet, at this level of approximation,
entanglement is instantaneous and no result about the dynamics of the decoherence
process can be extracted from the approximate reduced density matrix.

\n
Moreover notice that
 ${\mathcal I}(R,R)=1$ , ${\mathcal I}(R,R')=\overline{{\mathcal I}}(R',R)$
and $|{\mathcal I} (R,R')|\leq 1$. For   $N$ large  ${\mathcal I}(R,R')$ tends
to be exponentially close to zero for $R \neq R'$.

\n
 In (\cite{afft}) a concrete example was considered
in the case $N = 1$. The initial condition for the heavy particle were chosen
as a superposition of two identical wave packets heading one against the
other. The wave packet of an isolated heavy particle would have shown interference
fringes typical of a two slit experiment. The decrease in the interference
pattern, induced by the interaction with a light particle,  was computed
and taken as a measure of the decoherence effect.

\n
We want to give here a brief summary of the same analysis for any number
of light particles where the enhancement of the decoherence effect due to
multiple scattering is easily verified.

\n
Let the initial state be the coherent superposition of two wave packets
in the following form

\ba &&\phi(R) = b^{-1} \left( f^{+}_{\sigma}(R) +
f^{-}_{\sigma}(R) \right), ~ \  ~ b \equiv \| f^{+}_{\sigma} +f^{-}_{\sigma}
\|_{L^2} \label{2pack}
\\ &&f^{\pm}_{\sigma}(R)=\f{1}{\sigma^{3/2}}
f\left(\f{R \pm R_0}{\sigma}\right) e^{\pm i P_0 \cdot R},  \;\;\;\;\;
R_0,P_0 \in \erre^{3} \;\;\; \label{2pack2}
\ea

\n where $f $  is a real valued function in the Schwartz space     ${\mathcal S}(\erre^{3})$ with
$\|f\|_{L^2} =1$, $R_0 =(0,0,|R_0|)$, $P_0 =(0,0,-|P_0|)$.

\n
It is clear that under  the free evolution the two wave packets (\ref{2pack}) exhibit a significant  overlap and the typical interference effect is observed.

\n
On the othe hand, if we take into account the interaction with the light particles and introduce the further assumption  $\sigma \alpha \|\nabla V\|_{L^2}  \ll1$, it can be easily seen that $\rho^{a}(t )$ is approximated by

\begin{eqnarray}
\rho^{e} (t) & = & e^{-i t X_0} \rho^{e}_0 \, e^{i t X_0}
\end{eqnarray}
where $\rho_0^{e}$ has integral kernel
\begin{eqnarray}
& & \rho^{e}_0 (R, R')  =  \frac 1 {b^2} \left( \left| f^{+}_\sigma (R) \right|^2
+    \left| f^{-}_\sigma (R) \right|^2
 +  \Lambda    f^{+}_\sigma (R) {\overline
f^{-}_\sigma (R')} +    {\overline\Lambda}   f^{-}_\sigma (R)
{\overline f^{+}_\sigma (R')} \right)\label{rhoe}
\\
& & \Lambda  \equiv \prod_{j=1}^{N}  \left( \Omega_{+}(R_0)^{-1} \chi_j,
\Omega_{+}(-R_0)^{-1}  \chi_j \right)_{L^2} \label{lamb}
\end{eqnarray}

\n
The proof is easily obtained adapting the proof given in (\cite{afft}) for the case $N=1$.

\n
It is clear from (\ref{lamb}) that, if the interaction is absent, then
$\Lambda = 1$ and (\ref{rhoe}) describes the pure state corresponding to the
coherent superposition of $f^{+}_\sigma$ and $f^{-}_\sigma$ evolving
according to the free Hamiltonian.

\n
If the interaction with the light particles is present then $\Omega_{+}(R_0)^{-1} \neq I$ and   
$|\Lambda| \ll 1$ for $N$ large. For specific model interaction the factor $\Lambda$ can also be explicitely computed (see e.g. the one dimensional case treated in \cite{dft}).

\n
This means that the only
effect of the interaction on the heavy particle is to reduce the non diagonal terms in (\ref{rhoe})
by the factor $\Lambda$ and this means that
the interference effects for the heavy particle are correspondingly reduced.

\n
In this sense we can say that a (partial) decoherence effect on the heavy particle
has been induced and, moreover, the effect is completely characterized by
the parameter $\Lambda$.

\vs
\vs
\vs
\n
{\bf Acknowledgment.} We thank prof. G. Dell'Antonio for many stimulating discussions and a constant encouragement during the preparation of this paper.

\vs
\vs
\vs

\vs
\vs
\vs


\begin{thebibliography}{99}
\vs


\bibitem[AFFT]{afft} Adami R., Figari R., Finco D., Teta A.,  On the asymptotic behaviour of a quantum two-body system in the small mass ratio limit. {\em J. Physics A: Math. Gen.}, {\bf 37}, 7567-7580 (2004).

\bibitem[BGJKS]{bgjks} Blanchard Ph., Giulini D., Joos E., Kiefer C., Stamatescu
I.-O. eds., {\em Decoherence: Theoretical, Experimental and Conceptual
Problems}, Lect. Notes in Phys. 538, Springer, 2000.

\bibitem[CCF]{ccf} Cacciapuoti C., Carlone R., Figari R., Decoherence induced by scattering: a three dimensional model. {\em J. Phys. A: Math. Gen.}, {\bf 38}, no. 22, 4933-4946 (2005).

\bibitem[D]{d} Dell'Antonio G., Towards a theory of decoherence. {\em Int. J. Mod. Phys. B}, {\bf 18}, no. 4-5, 643-654 (2004).

\bibitem[DFT]{dft} D\"urr D., Figari R., Teta A., Decoherence in a two-particle model. 
{\em J. Math. Phys.} {\bf 45}, no. 4, 1291-1309 (2004).

\bibitem[DS]{ds} D\"urr D., Spohn H., Decoherence Through Coupling to the
    Radiation Field, in {\em Decoherence: Theoretical, Experimental
    and Conceptual Problems},  Blanchard Ph., Giulini D., Joos E.,
    Kiefer C., Stamatescu I.-O. eds., Lect. Notes in Phys. 538,
    Springer, 2000, pp. 77-86.

\bibitem[GJKKSZ]{gjkksz} Giulini D., Joos E., Kiefer C., Kupsch J., Stamatescu I.-O., Zeh
H.D., {\em Decoherence and the Appearance of a Classical World in
Quantum Theory}, Springer, 1996.

\bibitem[GF]{gf} Gallis M.R., Fleming G.N., Environmental and spontaneous localization. 
{\em Phys. Rev. A}, {\bf 42}, 38-48 (1990).

\bibitem[H]{h} Hagedorn G.A., A time dependent Born-Oppenheimer approximation.
{\em Comm. Math. Phys.}, {\bf 77},
n.1, 1--19 (1980).


\bibitem[HJ]{hj} Hagedorn G.A., Joye A., A time-dependent Born-Oppenheimer approximation
with exponentially small error estimates. {\em Comm. Math. Phys.}, {\bf 223},
n.3, 583-626 (2001).


\bibitem[HS]{hs} Hornberger K., Sipe J.E., Collisional decoherence reexamined. 
{\em Phys. Rev. A} {\bf 68}, 012105, 1--16 (2003).

\bibitem[HUBHAZ]{hubhaz} Hornberger K., Uttenhaler S., Brezger B., Hackerm\"uller L.,
Arndt M., Zeilinger A., Collisional decoherence observed in matter
wave interpherometry. {\em Phys. Rev. Lett.}, {\bf 90}, 160401 (2003).







\bibitem[JZ]{jz} Joos E., Zeh H.D., The emergence of classical properties through
interaction with the environment. {\em Z. Phys.} {\bf B59}, 223--243
(1985).
\bibitem[KY]{ky} Kato T.,  Yajima K., Some examples of smooth operators and the associated
smoothing effects, {\em Rev. in Math. Phys}, {\bf 1}, (1989), 481-496.

\n
\bibitem[RS]{rs} Rodiansky I., Schlag W.,
Time decay for solutions of Schr\"{o}dinger equations with rough and time-dependent potentials,
{\em Invent. Math.} {\bf 155} (2004), 451-513.

\bibitem[Sc]{s} Schlag W., Dispersive estimates for Schr\"{o}dinger operators: a survey, arXiv:math.AP/0501037.

\bibitem[Si]{si} Simon B., {\em Quantum Mechanics for Hamiltonians Defined as 
Quadratic Forms}, Princeton University Press, 1971.




\bibitem[Y]{y} Yajima, K., The $W\sp {k,p}$-continuity of wave operators
for Schr\"odinger operators. {\em J. Math. Soc. Japan} {\bf 47}, n.3,
551--581 (1995).
\end{thebibliography}
\end{document}